\documentclass{aa}
\usepackage{natbib,twoopt}
\usepackage[varg]{txfonts}
\usepackage{graphicx}

\bibpunct{(}{)}{;}{a}{}{,} 
\begin{document}
\title{The GTC exoplanet transit spectroscopy survey IV}
\subtitle{Confirmation of the flat transmission spectrum of HAT-P-32b}
\titlerunning{Confirmation of the flat transmission spectrum of HAT-P-32b}

\author{L.~Nortmann\inst{\ref{inst1}}
 \and E.~Pall\'e
 \inst{\ref{inst2},\ref{inst3}}
  \and F.~Murgas\inst{\ref{inst4},\ref{inst5}}
     \and S. Dreizler\inst{\ref{inst1}}
  \and N.~Iro\inst{\ref{inst6}}
  \and  A.~Cabrera-Lavers\inst{\ref{inst2},\ref{inst3}}
}

\institute{Institut f\"ur Astrophysik, Georg-August-Universit\"at G\"ottingen, Friedrich-Hund-Platz 1, D-37077 G\"ottingen, Germany \label{inst1} \email{nortmann@astro.physik.uni-goettingen.de}
\and Instituto de Astrof\'{\i}sica de Canarias (IAC), 38205, La Laguna, Tenerife, Spain \label{inst2} 
\and Departamento de Astrof\'{\i}sica, Universidad de La Laguna (ULL), 38206, La Laguna, Tenerife, Spain\label{inst3}
\and Univ. Grenoble Alpes, IPAG, F-38000 Grenoble, France\label{inst4}
\and  CNRS, IPAG, F-38000 Grenoble, France\label{inst5}
\and Theoretical Meteorology group, Klimacampus, University of Hamburg, Grindelberg 5, 20144 Hamburg, Germany\label{inst6}}

\date{Received 8 September 2015;
Accepted 19 April 2016}

\abstract {We observed the hot Jupiter HAT-P-32b (also known as HAT-P-32Ab) to determine its optical transmission spectrum by measuring the wavelength-dependent planet-to-star radius ratios in the region between 518 - 918 nm. We used the OSIRIS instrument at the GTC in long slit spectroscopy mode, placing HAT-P-32 and a reference star in  the same slit and obtaining a time series of spectra covering two transit events. 
Using the best quality data set, we were able to yield 20 narrow-band transit light
curves, with each passband spanning a 20 nm wide interval. 
After removal of all systematic noise signals and light curve modeling the uncertainties for the resulting radius ratios lie between 337 and 972 ppm. The radius ratios show little variation with wavelength suggesting a high altitude cloud layer masking any atmospheric features. Alternatively, a strong depletion in alkali metals or a much smaller than expected planetary atmospheric scale height could be responsible for the lack of atmospheric features. Our result of a flat transmission spectrum is consistent with a previous ground-based study of the optical spectrum of this planet. This agreement between independent results demonstrates that ground-based measurements of exoplanet atmospheres can give reliable and reproducible results despite the fact that the data often is heavily affected by systematic noise, as long as the noise source is well understood and properly corrected.\newline
We also extract an optical spectrum of the M-dwarf companion HAT-P-32B. Using PHOENIX stellar atmosphere models we determine an effective temperature of $T_\mathrm{eff}= 3187^{+60}_{-71}$ K, slightly colder than previous studies relying only on broadband infra-red data.} 

\maketitle
\section{Introduction}\label{chap:int}
Two decades after the first detection of an exoplanet around a solar-type star by \citet{1995Natur.378..355M} the field of exoplanet science is fast-moving and has expanded into many sub-fields. A new main focus is the characterization of exoplanet atmospheres. The most successful approach to studying the planet's atmospheric properties has been the measurement of their transmission and emission spectra from multi-color observations of the occultation events in transiting planetary systems. While the emission of the planet can be inferred from the drop in flux during the secondary eclipse the planet's transmission spectrum can be obtained during the primary eclipse. This is possible since the planet's atmosphere will be opaque at wavelengths where the atmospheric constituents absorb light causing a larger effective planet radius and, thus, a deeper transit. Many successful measurements of wavelength-dependent planet radii have been obtained from space using the Hubble Space Telescope (HST)  \citep[e.g.][]{2002ApJ...568..377C,2008MNRAS.385..109P, 2012ApJ...747...35B,2015MNRAS.446.2428S}. Moreover, in the last 4 years, ground-based observations have also yielded promising results  \citep[e.g.][]{2010Natur.468..669B,2014A&A...563A..41M,2013ApJ...778..184J,2013MNRAS.428.3680G}. However, both space-based and ground-based data often are affected by systematic noise signals, which need to be addressed before a high quality transmission spectrum can be extracted. In the past, the correct treatment of these noise signals has been subject of scientific debate and has led to disagreements between the conclusions of several groups studying the same data sets \citep[e.g.][for HD 189733\,b]{2007Natur.448..169T,2007ApJ...668L.179E,2009ApJ...699..478D,2011MNRAS.411.2199G}. As a general consequence this has created doubts concerning the robustness of presented results. In this paper we aim to demonstrate on the case of HAT-P-32b that reliable results for a planet's transmission spectrum can be obtained from the ground. The hot Jupiter with a mass of $M=0.860 \pm 0.16$ $M_\mathrm{Jup}$ and a radius of $R= 1.789 \pm 0.025$ $R_\mathrm{Jup}$ was discovered by \citet{2011ApJ...742...59H} around an late-type F dwarf star (Vmag=11.44) at an  $2.15$ day orbit. The planet's dayside temperature was measured to be $T_\mathrm{eq}=2042 \pm 50$ K by \citet{2014ApJ...796..115Z} from  secondary eclipse observations in the $H$, $K_S$, $3.6$ and $4.5$ $\mathrm{\mu m}$ bands. An optically close companion was discovered in \citeyear{2013AJ....146....9A} by \citeauthor{2013AJ....146....9A} The stellar companion was recently studied in more detail and concluded to be an M-dwarf bound to the HAT-P-32 system from proper motion and AO measurements \citep{2015ApJ...800..138N,2014ApJ...796..115Z}. 
Both studies place the effective temperature of the companion at about $T_\mathrm{eff}\approx3500$ K. Following the notation used in these works in the following we will  refer to this stellar companion as HAT-P-32B and to the planet host star and the planet as HAT-P-32A and HAT-P-32Ab, respectively. 
\citet{2014ApJ...785..126K} observed the HAT-P-32 system among several other planet host stars for radial velocity (RV) trends that could indicate additional companions. They found a long trend signal for HAT-P-32A pointing to the existence of yet another body in the system. A transit timing variation (TTV) study of 45 transit events by \citet{2014MNRAS.441..304S} looking for evidence of an additional body found no evidence for variations larger than 1.5 min.\newline
\citet{2013MNRAS.436.2974G} obtained a ground-based optical transmission spectrum of HAT-P-32Ab using Gemini North/GMOS. Their results show a flat transmission spectrum. \newline
In the study presented in this paper we probed a very similar wavelength range using the long slit method at the OSIRIS instrument at the 10-meter class telescope GTC aiming to verify the nature of the transmission spectrum and further demonstrate the potential of GTC/OSIRIS as a reliable survey instrument for observations of this kind.\newline  
The paper is organized as follows. We first will describe the observing set up and data reduction in Sect. \ref{sec:obs}. This is followed by a description of the white light curve analysis in Sect. \ref{sec:wlfit} and a discussion of the white light curve results in Sect. \ref{sec:resdiswhite}. Here we also address systematic noise signals we found in both data sets. In Sect. \ref{sec:systematics} we will describe the source of the largest noise signal and motivate its correction for the narrow band light curves followed by a description of the extraction of the transmission spectrum during the analysis of the narrow band light curves in Sect \ref{sec:colorfit}. We present and discuss our results for the transmission spectrum in Sect. \ref{sec:results} and draw our conclusions in Sect. \ref{sec:conclusion}. The study of photometric and spectroscopic data of the companion HAT-P-32B in order to derive its stellar parameters and extract values necessary for the correction of its diluting effect on the transit depth of HAT-P-32Ab can be found in the Appendix \ref{sec:sec}.

\section{Observations and data}\label{sec:obs}
\begin{table}
\begin{center}
\setlength{\tabcolsep}{1.5mm}
\begin{tabular}{c c c c}
\hline\hline
Star&RA&DEC& Vmag\\ \hline
\hspace{20mm}\newline
HAT-P-32    &    	 $02$h $04$m $10.278$s&$+46\degr$ $41\arcmin$ $16.21\arcsec$& 11.44 \\ 
Ref1 (Run 1) &   $02$h $04$m $15.060$s&$+46\degr$ $40\arcmin$ $49.57\arcsec$& 13.59 \\
Ref2 (Run 2)   &    $02$h $03$m $51.771$s&$+46\degr$ $41\arcmin$ $32.23\arcsec$ &10.97\\
 \hline
\end{tabular}
\caption{Coordinates of the planet host star HAT-P-32A and the reference stars `Ref1' and `Ref2' used in the first and second observing run.}  
\label{tab:coord}  
\end{center}
\end{table}
We observed HAT-P-32Ab twice during transit on 2012  September 15 (JD 2456185.5, hereafter referred to as \textit{Run 1}) and on 2012 September 30 (JD 2456200.5, hereafter referred to as \textit{Run 2}) with the OSIRIS instrument (Optical System for Imaging and low-Intermediate-Resolution Integrated Spectroscopy; \citet{2012SPIE.8446E..4TS}) mounted at the Spanish 10.4 m telescope GranTeCan (GTC). We chose the method of long slit spectroscopy, in which the planet host star and a suitable reference star are both placed inside one long slit. The grism R1000R was used to disperse the light over the range from 518 to 918 nm. With an exposure time of 10 seconds (Run 1) and 7 seconds (Run 2) continuous time series of 321 (Run 1) and 700 (Run 2) optical spectra were obtained in each night covering the duration of the whole transit event in both cases. We used slightly different observing set-ups in each night, working with different regions of the CCD detector. The OSIRIS detector consists of a mosaic of two $2048\times4096$ pixel Marconi CCD42-82 chips. 
\begin{figure}
    \resizebox{\hsize}{!}{\includegraphics{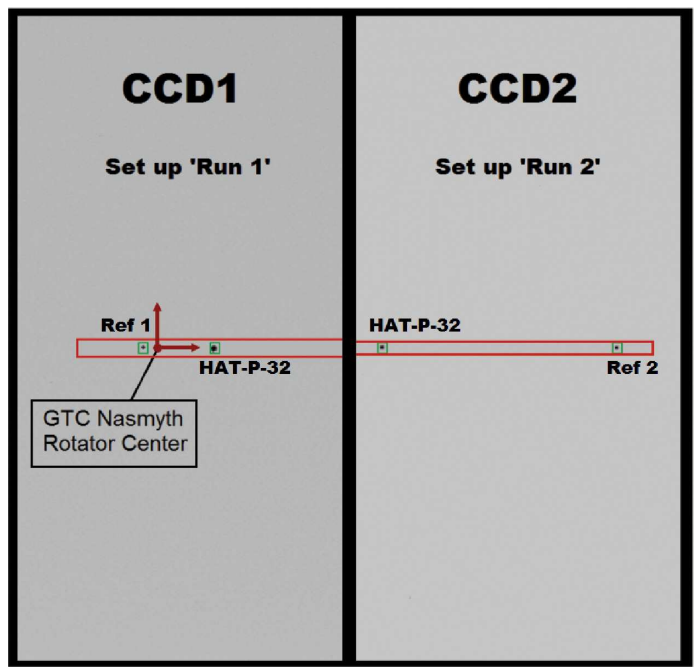}}	
  	\caption{Observing set up for Run 1 and Run 2. In Run 1 both stars were placed in CCD1 and a $12\arcsec$ wide slit was used. In Run 2 we placed both stars in CCD2 and used a slit of $10\arcsec$ width.}
	\label{fig:setup}
\end{figure}
In Run 1 the chosen reference star `Ref1' was considerably fainter ($\Delta$ Vmag $=2.15$) than HAT-P-32A and located at a distance of $56.0\arcsec=0.93\arcmin$. A custom made $12\arcsec$ wide slit was used and both stars were placed on CCD1. In Run 2 we chose a brighter reference star ($\Delta$ Vmag $=-0.467$) located at a $191.0\arcsec=3.18\arcmin$ separation from HAT-P-32A. Due to the larger distance between the stars, they could not both be placed in CCD1. In order to still have both stars on the same CCD and, thus, avoid possible complications from variations in the detector properties we placed both stars on CCD2 which has a larger unvignetted field of view through the slit than CCD1. We also exchanged the custom made $12\arcsec$ wide slit for a $10\arcsec$ wide slit, since the latter is slightly longer (extending $0.567\arcmin$ further into CCD2) giving both stars more room in spatial direction. The set up for both observing runs is illustrated in Fig. \ref{fig:setup} and the coordinates for both reference stars are given in Table \ref{tab:coord}. The observing conditions during both nights were good, with an average seeing of $1.06\arcsec$ in Run 1 and $1.12\arcsec$ in Run 2. The seeing was not stable in either run, varying between $0.78\arcsec$ and $1.82\arcsec$ during the first, and between $0.69\arcsec$ and $2.42\arcsec$ during the second night.\newline 
Due to complications during the observation almost no out-of-transit data was obtained during Run 1 and part of the data was rendered useless by a light reflection passing over the detector contaminating the red part of the spectrum of HAT-P-32A for approximately 20 minutes (35 frames) during the second half of the transit. An example of a contaminated frame is shown in Fig. \ref{fig:reflec}. 
\begin{figure}
  \resizebox{\hsize}{!}{\includegraphics{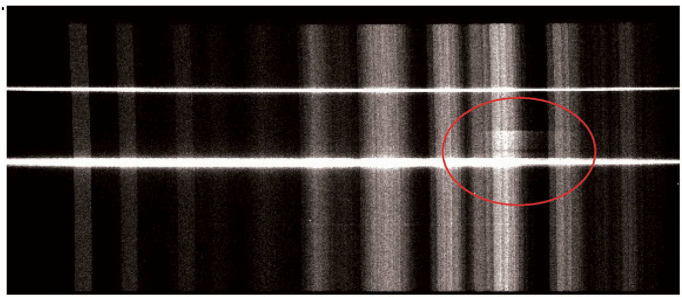}}	
  	\caption{Example of an image obtained in Run 1 which was contaminated by an internal light reflection. The image shows the light of HAT-P-32A (bottom) and the fainter reference star Ref1 (top) dispersed by the grism in horizontally direction. The contaminating light reflection is marked by a red circle.}
	\label{fig:reflec}
\end{figure}
After each run about 50 bias frames and 100 sky flats were taken. For wavelength calibration, spectra of the HgAr, Xe and Ne lamps were taken with a $1\arcsec$ wide slit.

\subsection{Data reduction and spectral extraction}\label{sec:specextract}
We employed standard data reduction procedures, subtracting the median averaged bias and dividing by the median averaged flat field. We then applied a wavelength calibration to every column of the image re-binning the data to a homogeneous wavelength grid using the IDL routine \texttt{rebinw} from the PINTofALE package \citep{2000HEAD....5.2705K}, guaranteeing flux conservation. This step ensured that every pixel in an image row corresponded to the same wavelength. The extraction of the stellar spectra from the images was then performed using the optimal extraction algorithm \citep{1986PASP...98..609H}. The algorithm performed well on our data sets, yielding lower noise levels than other, more simple approaches. During the extraction a median averaged spatial profile was used to identify and mask cosmic ray strikes. Due to the close projected distance of HAT-P-32A and the fainter M-dwarf companion (which was not yet discovered at the time of the observations) it was not possible to reliably exclude HAT-P-32B by choosing a narrow extraction aperture. Instead we choose a very wide aperture (80 pixel = 20.32\arcsec) ensuring that HAT-P-32B was fully within this aperture at all times. As a consequence the diluting effect of its additional flux on the transit depth had to be corrected during transit modeling (see Sect. \ref{sec:wlfit}).\newline
During the observations the stars drifted slightly in spatial and also in dispersion direction. We monitored the drift in spatial direction by fitting a Gaussian function to the stellar profile tracing the position of the peaks. In this step we also retrieved the FWHM of the fitted Gaussian profile to monitor the seeing variations. The drift of the stars in dispersion direction caused small shifts in the wavelength solution with respect to the one obtained for a star perfectly centered within the slit. We monitored and corrected these wavelength shifts by calculating the cross correlation of each spectrum with the first spectrum of the respective run. 

\subsection{Light curves}\label{sec:lightcurves}
\begin{figure}
    \resizebox{\hsize}{!}{\includegraphics{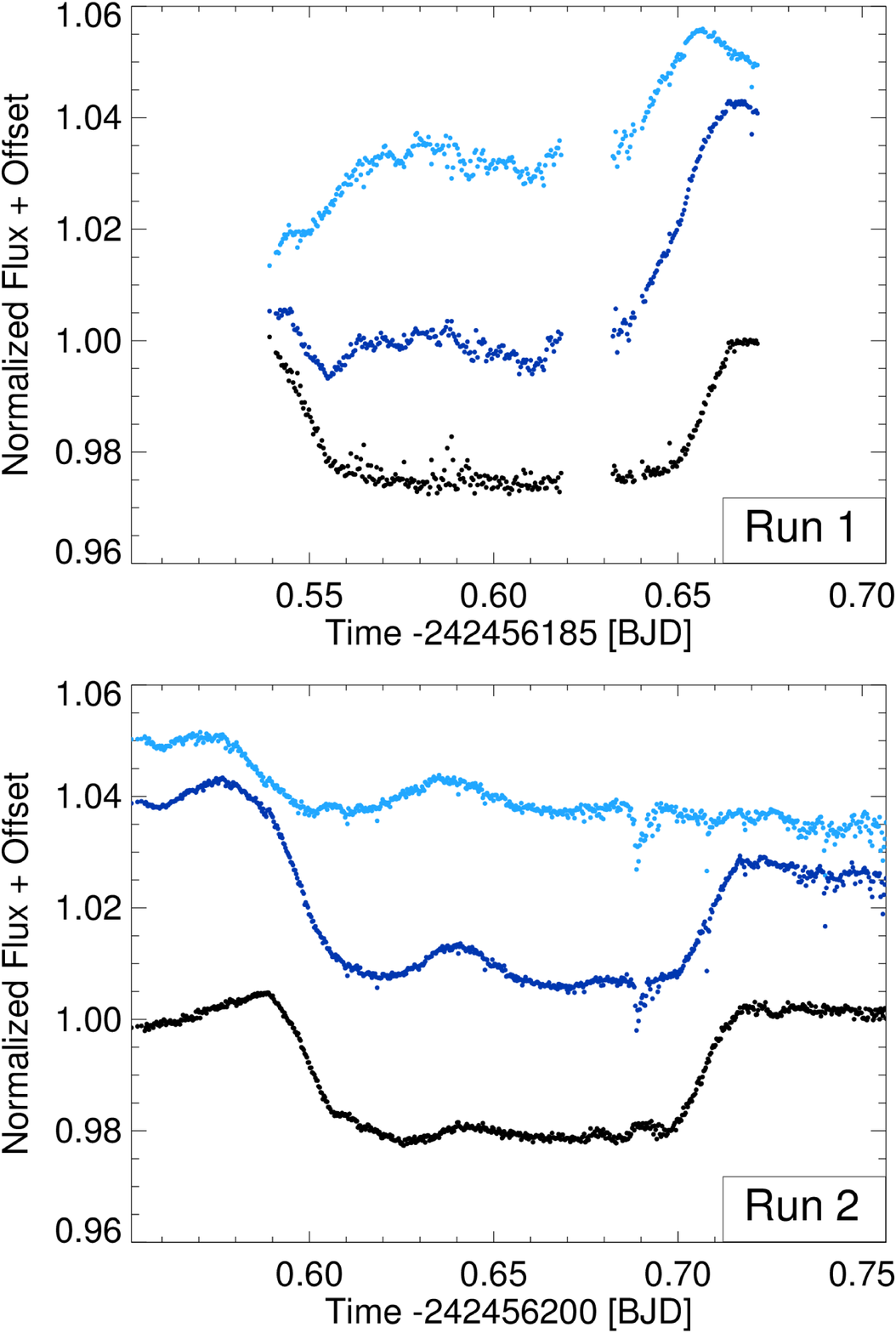}}	
    \caption{Raw white light curves for HAT-P-32A (dark grey circles) and the reference star (light grey circles) for both observing runs and their division the relative white light curve (black circles). \textit{Top panel:} data for Run 1, \textit{bottom panel:} data for Run 2.}
    	\label{fig:rawwlboth}
    \end{figure}

\begin{figure}
  \resizebox{\hsize}{!}{\includegraphics{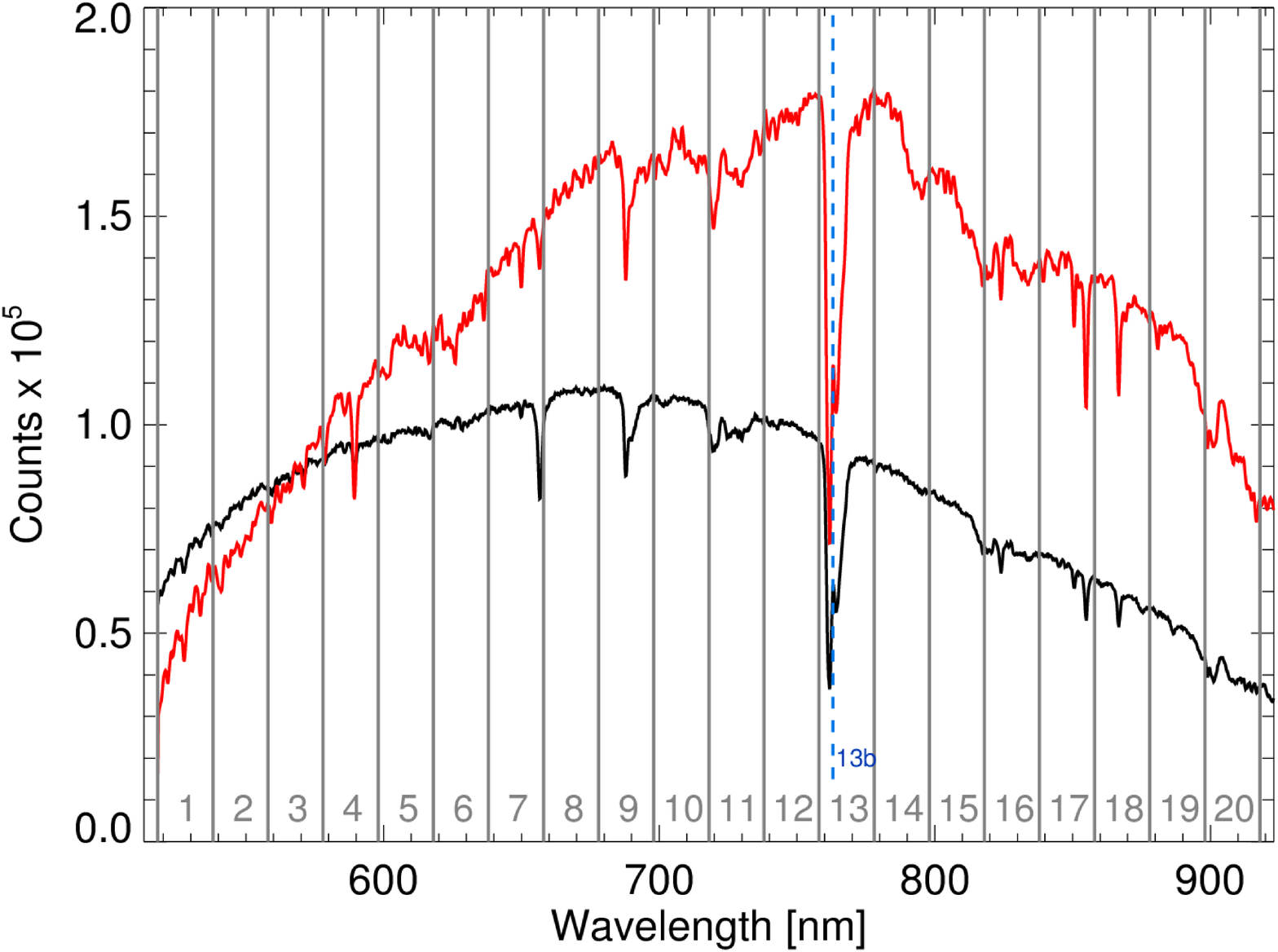}}
  	\caption{Spectra of HAT-P-32A and reference star Ref2 from Run 2. Indicated in grey are the limits of the twenty 20 nm wide narrow band channels. The blue limit of the additionally defined 15 nm wide channel  \#13b excluding the first 5 nm of channel \#13 is indicated with a blue dashed line.}
	\label{fig:spectra_run2}
\end{figure}

\begin{figure}
  \resizebox{\hsize}{!}{\includegraphics{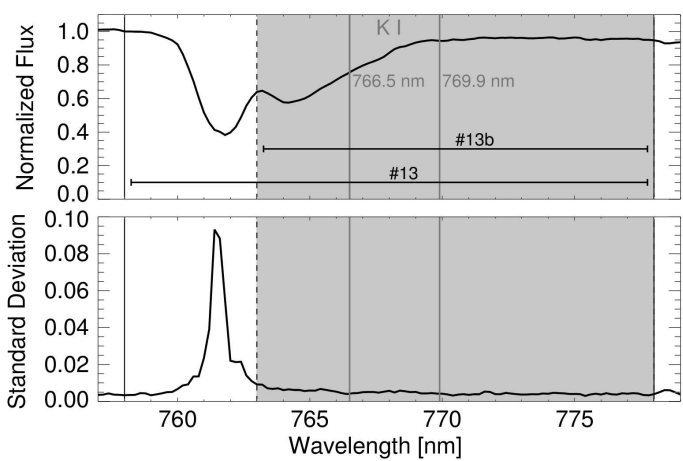}}
  	\caption{Definition of narrow band channel \#13b as a sub channel of channel \#13. The grey shaded area encompasses the predicted \ion{K}{I} resonance lines but avoids the high noise region caused by telluric oxygen absorption. \textit{Top panel:} Example spectrum of Ref2 within the limits of narrow band channel \#13, showing the significant flux decrease in the telluric oxygen. \textit{Bottom panel:} Noise level at each wavelength as estimated from the after transit light curve standard deviation showing a strong increase of noise in the stronger telluric oxygen absorption band.}
	\label{fig:c13b}
\end{figure}

Since any telluric variations during the observing runs will have affected both planet host star and reference star in the same manner, these effects can be neglected if only the relative light curves of these two objects are considered. We created white light curves from the data by dividing the total summed flux over all wavelengths for HAT-P-32A by the total sum of the reference star spectra for each measurement. The resulting white light curves are shown in Fig. \ref{fig:rawwlboth}. We further created twenty narrow band channel light curves from the data of Run 2 by dividing the wavelength range into intervals of 20 nm.  The spectra of HAT-P-32A and reference star Ref2 are shown in Fig. \ref{fig:spectra_run2} together with the channel limits. Channel \#13 encompassed both the telluric oxygen bands and the potassium resonance lines (\ion{K}{I} at $766.5$ and $769.9$ nm) predicted for exoplanet atmospheres at moderate temperatures. Akin to what has been reported by \citet{2016A&A...585A.114P} we found that the noise level, estimated from the standard deviation of the out-of-transit light curve scatter in each individual wavelength point, rises significantly in the deeper of the two telluric oxygen bands, negatively affecting the signal-to-noise of the entire channel \#13 light curve.  We therefore constructed an additional channel (channel \#13b) of only $15$ nm width, spanning from $763$ to $788$ nm. This channel was still encompassing the expected potassium lines but avoided the high noise region as sketched in Fig. \ref{fig:c13b}.

\section{Analysis of the white light curves}\label{sec:wlfit}
\begin{figure}
  \resizebox{\hsize}{!}{\includegraphics{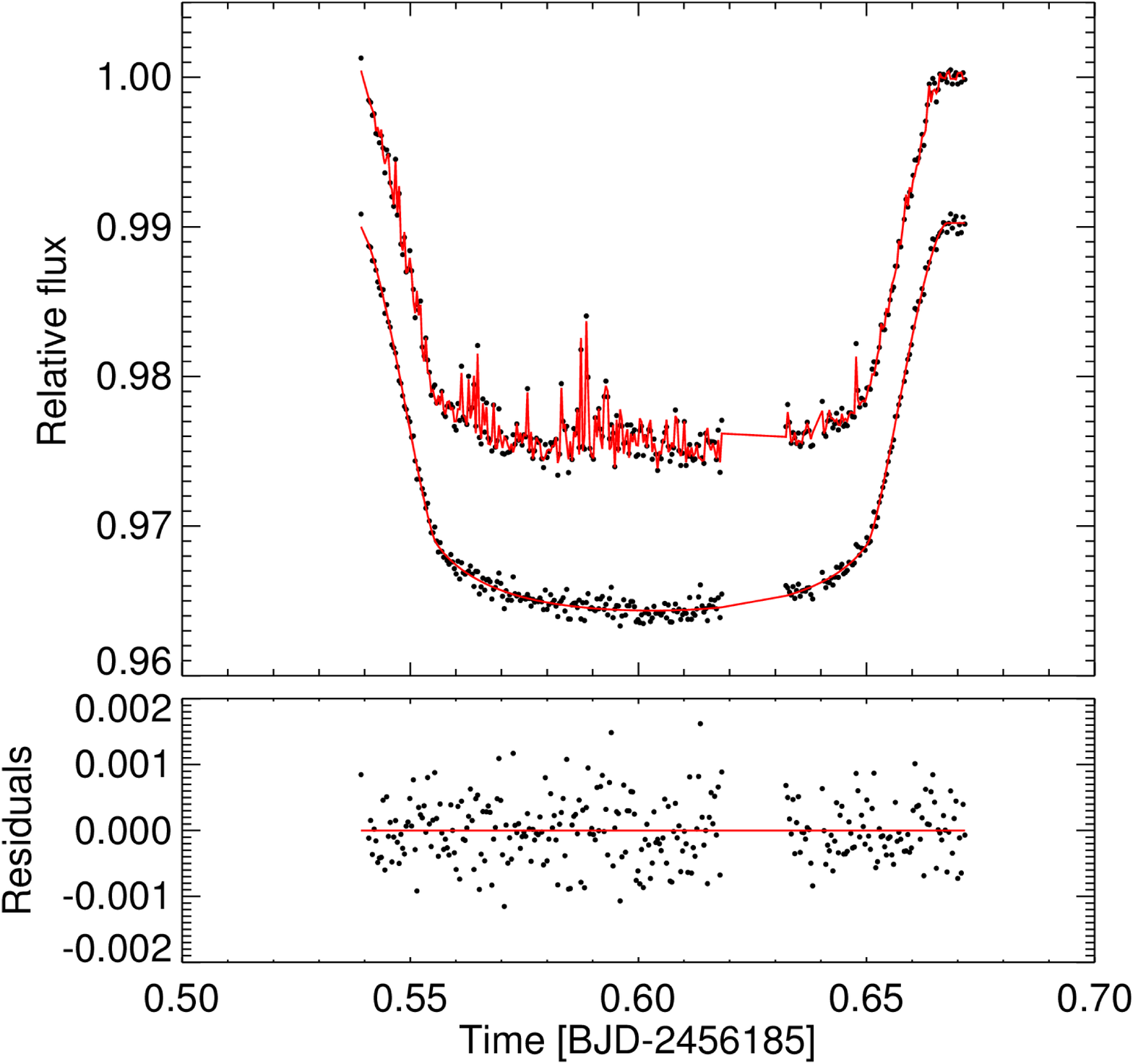}}
  	\caption{\textit{Top panel:} White light curve from Run 1 (black points) with best fit model (red line) before (top) and after (bottom) division by the best fit systematic noise model (see Sect. \ref{sec:wlfit}). The bottom curve was shifted downwards by a small offset for clarity. \textit{Bottom panel:} Residuals of the white light curve from Run 1 after subtraction of the best fit model.}
	\label{fig:white_fit_r1}
\end{figure}

The relative white light curves for both runs are shown in Fig. \ref{fig:rawwlboth}. We found both white light curves to be affected by red noise. Previous works dealing with data obtained with GTC/OSIRIS have reported on similar noise signals \citep{2012MNRAS.426.1663S,2014A&A...563A..41M,2015A&A...580A..60M}. The three works explored different systematics models which included terms depending on the seeing and air mass. We also found indicators for a possible correlation of the red noise with auxiliary parameters and explored these possible correlations by including a systematic noise model in our light curve fitting. We found the minimal $\chi^2$ using the IDL implementation of the Levenberg-Markward algorithm \texttt{mpfit} by \citet{2009ASPC..411..251M} while modeling the light curves with a model of the form:
\begin{equation}
	\mathcal{M}=\left(\mathcal{T}+c_w\right)\cdot\mathcal{S}
\end{equation}
where $\mathcal{T}$ is the analytical transit model described by \citet{2002ApJ...580L.171M}, $\mathcal{S}$ is a systematic noise model and $c_w$ is the relative flux contribution of the stellar companion HAT-P-32B in white light (i.e. $c_w=f_{\text{HAT-P-32B}}/f_{\text{HAT-P-32A}}$, where $f$ is the total flux integrated over the white light wavelength range). In the following we will refer to $c_w$ and its equivalents for the narrow band channels as the dilution factor. 
The \citeauthor{2002ApJ...580L.171M} transit model parameterizes the transit using the radius ratio between the planet and the star
$R_p/R_{\star}$, the semi-major axis of the planet's orbit in units of the
stellar radius $a_p/R_{\star}$, the inclination of the planet orbit $i$, a reference
time for the mid-transit $T_\mathrm{C_1}$ and two coefficients $u_1$ and $u_2$ describing the stellar limb darkening with a quadratic limb darkening
law. Also needed is the period of the planet's orbit $P$, which we kept fixed to the value given by \citet{2011ApJ...742...59H} ($P = 2.150008$ days) and the eccentricity of the orbit $e$, which we kept fixed at $e = 0$. The code implementation of \citet{2013PASP..125...83E} was used for the calculation of the transit model. The relative flux contribution of the stellar M-dwarf companion $c_w$ and its uncertainty were determined from our own data of Run 2 as described in detail in Appendix \ref{sec:sec} and this value was kept fixed during this optimization step. We tested 135 different systematic noise models $\mathcal{S}$ where each model was a different combination of polynomial functions depending on the four auxiliary parameters: position in spatial direction $xpos$, position in dispersion direction $ypos$, seeing/FWHM of the stellar profile $fwhm$ and air mass $airm$. All combinations of different polynomial orders between 0 and 2 (between 0 and 4 for the FWHM) for all four auxiliary parameters were explored. The most complex form of $\mathcal{S}$ we tested, consequently, was of the form:
\begin{equation}
\begin{split}
	\mathcal{S}=~&n_0+\\
& 	x_1\cdot xpos+x_2\cdot xpos^2+\\
&	y_1\cdot ypos+y_2\cdot ypos^2+\\
& 	a_1\cdot airm+a_2\cdot airm^2+\\
&	f_1\cdot fwhm+f_2\cdot fwhm^2+f_3\cdot fwhm^3+f_4\cdot fwhm^4
	\end{split}
\end{equation}
where $n_0$ is the normalization and $x_{1,2}$, $y_{1,2}$, $a_{1,2}$ and $f_{1,2,3,4}$ are model parameters scaling the influence of the respective auxiliary parameter.\newline
In order to avoid over fitting due to the fact that the minimal $\chi^2$ generally decreases with a rising number of free parameters, a compromise between the number of free parameters and minimization of the $\chi^2$ should be found by the means of an objective criterion. We used the Bayesian Information Criterion for the case of an unknown variance of the data
\begin{equation}
BIC =N\ln(RSS/N) + k \cdot \ln(N)
	\label{eq:bic}
\end{equation}
where $RSS$ is the sum of squared residuals, $k$ is the number of free parameters and $N$ is the number of data points, to determine the best choice of $\mathcal{S}$.
We compared the $BIC$ of all tested models and found that using a combination of a first-order polynomial of the position drift in dispersion direction ($ypos$), a first-order polynomial of the air mass and a third-order polynomial function of the FWHM yields the lowest BIC for the data of Run 1, i.e. a model of the form: 
\begin{equation}
\begin{split}
	\mathcal{S}_{\mathrm{Run 1}}^*=~&n_0+y_1\cdot ypos +	a_1\cdot airm+\\
&	f_1\cdot fwhm+f_2\cdot fwhm^2+f_3\cdot fwhm^3
	\end{split}
	\label{eq:syswr1}
\end{equation}
An abbreviated compilation of the results of the $BIC$ comparison can be found in Table \ref{tab:bich321}, where, for practicality, only the models with an $\Delta BIC=BIC- BIC_{\text{min}} <10$ are listed. The best fit results for the planet-to-star radius ratio derived by using these model approaches differ only slightly, and all lie well within the $1\sigma$ uncertainty interval (derived in Sect. \ref{sec:wlerror}) of the best fit result obtained with the lowest BIC model. In Table \ref{tab:bich321} the models are identified by a short code, which is to be read as follows:
`xpos' is the keyword for the position drift in spatial direction, `ypos' for the drift in dispersion direction, `airm' for the air mass and `fwhm' for the seeing/FWHM of the stellar profile. The number following each of these keywords indicates the highest polynomial order that was allowed to be non-zero and free in the model fit. Using the systematic noise model $\mathcal{S}_{\mathrm{Run 1}}^*$ given in Eq. \ref{eq:syswr1} we achieved a good fit to the data of Run 1 with an almost Gaussian distribution of the residuals (with a normalized standard deviation of $463$ ppm). The remaining correlation of the residuals is explored in the Sect. \ref{sec:wlerror}, which focusses on the error estimates of the results. A plot of the white light curve with the best fit model can be found in Fig. \ref{fig:white_fit_r1}.
 \begin{table}
\begin{center}
\begin{tabular}{c c c}
\hline\hline
Model &$\Delta {BIC}$ &  $\Delta R_p/R_\star$
\\ \hline
xpos0 ypos0 airm0 fwhm3 &        5.76           &  ~0.000586\\
xpos0 ypos0 airm1 fwhm3 &        4.28  					&  ~0.000292\\
xpos0 ypos0 airm1 fwhm4 &        9.63  					&  ~0.000332\\
xpos0 ypos1 airm0 fwhm3 &        7.96  					&  ~0.000399\\
xpos0 ypos1 airm1 fwhm2 &        6.92 					&  ~0.000050\\
xpos0 ypos1 airm1 fwhm3 &        \textbf{0.00}  &  ~0.000000\\
xpos0 ypos1 airm1 fwhm4 &        5.70 					&  -0.000125\\
xpos0 ypos2 airm1 fwhm3 &        4.63  					&  ~0.000019\\
xpos1 ypos0 airm0 fwhm3 &        7.23  					&  ~0.000300\\
xpos1 ypos0 airm1 fwhm3 &        9.77  					&  ~0.000443\\
xpos1 ypos0 airm2 fwhm3 &        7.46  					&  ~0.000492\\
xpos1 ypos1 airm0 fwhm3 &        5.54  					&  -0.000456\\
xpos1 ypos1 airm1 fwhm3 &        5.69  					&  ~0.000046\\
xpos1 ypos2 airm0 fwhm3 &        9.77  					&  -0.000346\\ \hline
\end{tabular}
\caption{Model comparison for the white light curve of Run 1. Different model approaches for the systematic noise and the respective values for $\Delta  {BIC}= {BIC}- {BIC}_{ \mathrm{min}}$ and the respective change in best fitting planet-to-star radius ratio $\Delta R_p/R_\star= R_p/R_\star- \left(R_p/R_\star\right)_{{BIC}_{ \mathrm{min}}}$.}  \label{tab:bich321}
\end{center}
\end{table}
\begin{figure}
  \resizebox{\hsize}{!}{\includegraphics{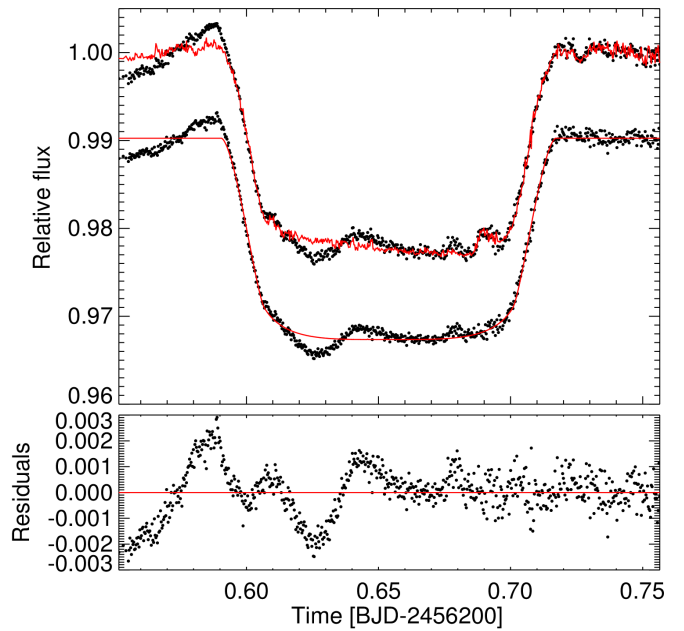}}
  	\caption{\textit{Top panel:} White light curve from Run 2 (black points) with best fit model (red line) before (top) and after (bottom) division by the best fit systematic noise model (see Sect. \ref{sec:wlfit}). The bottom curve was shifted downwards by a small offset for clarity. \textit{Bottom panel:} Residuals of the white light curve from Run 2 after subtraction of the best fit model.}
	\label{fig:white_fit_r2}
\end{figure}
After repeating this process for the data of Run 2 we found the minimal $BIC$ for a systematics model consisting of a first-order polynomial of the position drift in spatial direction ($xpos$), a first-order polynomial of the air mass and a third-order polynomial function of the FWHM. However, neither this model nor any other tested combination did yield a good fit to the whole curve leaving the residuals still strongly correlated (see Fig. \ref{fig:white_fit_r2}).
This lead us to conclude that the systematic noise, which is strongly distorting the first part of the light curve of Run 2, has a different origin. We will investigate this origin in Sect. \ref{sec:systematics}.

\subsection{Errors}\label{sec:wlerror}
\begin{figure*}
  \resizebox{\hsize}{!}{\includegraphics{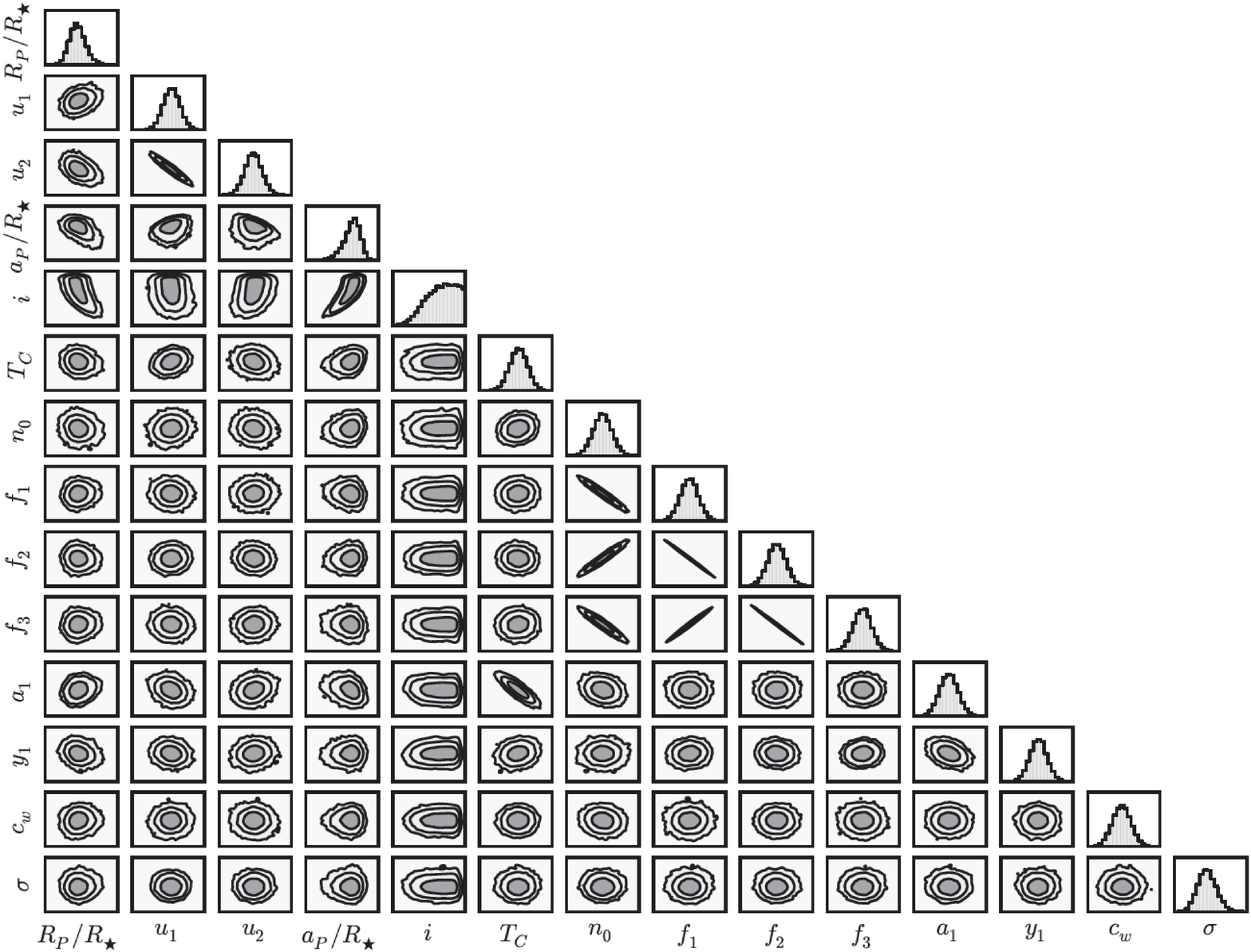}}	
  	\caption{Correlation plots of the posterior parameter distributions for the white light curve of Run 1 from MCMC. The lines indicate the areas encompassing $1\sigma$, $2\sigma$ and $3\sigma$ of the whole distribution, respectively.
  	}
	\label{fig:confidence_intervals_h321_wlight}
\end{figure*}

For the white light curve fit of Run 1 we probe the probability distributions of the model parameters with  Markov Chain Monte Carlo (MCMC) sampling using the \texttt{emcee} package \citep{2013PASP..125..306F}. 
We used non-informative priors for all parameters except the dilution coefficient $c_w$, which we allowed to vary within its uncertainties using a Gaussian prior. 
The MCMC sampling was run with an ensemble of 600 walkers  each starting with slightly different parameter populations as seeds for the chains. We let the chains run for 20\,000 accepted steps and defined a burn-in phase, in which the chains are not yet fully converged, to be over after 5000 steps omitting all chain-steps prior to this mark. The results can be considered robust if all the chains converge to the same probability distribution, which they did in our case. We determined a thinning factor for each chain as the largest autocorrelation length of any parameter within the chain. The average thinning factor was $\approx 200$ which resulted in a total of 45\,891 accepted sample points for the merged distribution of an accepted chain steps. The correlation plots of the posterior distributions for all parameters of the white light curve fit of Run 1 are shown in Fig. \ref{fig:confidence_intervals_h321_wlight}.\newline
The $1\sigma$ uncertainties of each parameter were calculated as the limits encompassing  $68.27\%$ of all sampled points.

\subsubsection{Red noise estimation}
\label{sec:rednoise}
In order to determine the level of red noise remaining in the light curve residuals and consequent underestimation of the parameter uncertainties, we calculated the red noise factor $\beta$. This factor was introduced by \citet{2008ApJ...683.1076W} and is based on a comparison of the progression of the standard deviation of the time-binned light curve residuals with the behavior expected for a light curve purely affected by Gaussian (i.e. white) noise. The latter is described by Eq. \ref{eq:sigmabeta} if $\sigma_1$ is the amplitude of the Gaussian white noise, $N$ the number of adjacent points binned together and $M$ the total number of bins.
\begin{equation}
	\sigma_N^\mathrm{theory}=\frac{\sigma_1}{\sqrt{N}}\sqrt{\frac{M}{M-1}}
	\label{eq:sigmabeta}
\end{equation}
The difference between theoretical and actual progression of the standard deviation of the binned residuals is caused by red noise. \citeauthor{2008ApJ...683.1076W} define the ratio between the two curves as the `red noise factor' $\beta=\sigma_N^\mathrm{actual}/\sigma_N^\mathrm{theory}$ which can be used to inflate the MCMC chain derived error bars of all parameters. We determined the maximal value of this ratio to be $\beta_{\mathrm{Run 1}}=1.345$ for the white light curve data from the first run when the systematics model with the lowest $BIC$ was used and $\beta_\mathrm{Run 2}\geq6.5$ for the data from Run 2, regardless of which systematics model $\mathcal{S}$ was used. The $\beta$ factor close to unity for Run 1 demonstrates that the systematics model $\mathcal{S}_{\mathrm{Run 1}}^*$ (Eq. \ref{eq:syswr1}) used was appropriate to correct all noise sources. In contrast, the high value derived for Run 2 clearly indicates a remaining noise source (see Sect. \ref{sec:systematics}), suggesting the model approach used for Run 1 to be insufficient for the analysis of the data of Run 2.

\section{Results and discussion of the white light curves}\label{sec:resdiswhite}
\begin{table*} 
\begin{center}
\def\tol#1#2#3{\hbox{\rule{0pt}{15pt}${#1}^{{#2}}_{{#3}}$}}
\begin{tabular}{cccccc}
\hline\hline 
Parameter &This work & \citet{2011ApJ...742...59H} ($e\equiv0$)  & \citet{2013MNRAS.436.2974G} & 
\citet{2014MNRAS.441..304S}\\
Filter/Range&$518-918$ nm &$I$, $z$, $g$ &$520.6-932.5$ nm & $r_\mathrm{S}$, $R_\mathrm{B}$,  $R_\mathrm{C}$, Clear \\  \hline
 
$R_\mathrm{p}/R_{\star}$ &
\tol{0.1516376}{+0.000874}{-0.000545}  &
\tol{0.1508}{+0.0004}{-0.0004}&
\tol{0.1515}{+0.0012}{-0.0012}&
\tol{0.1510}{+0.0004}{-0.0004}\\

$a_\mathrm{p}/R_{\star}$ &
\tol{6.123}{+ 0.021}{-0.054} &
 \tol{6.05}{+0.03}{-0.04}&
 \tol{6.091}{+0.036}{-0.047}&
 \tol{6.056}{+0.009}{-0.009} \\ 
 
$i$ 						&
\tol{89.33}{+0.58}{-0.80} &
\tol{88.9}{+0.4}{-0.4}&
\tol{89.12}{+0.61}{0.68} &
\tol{88.92}{+0.10}{-0.10}\\

$P$ [d] 		  	&
\tol{2.1500080}{}{} (fixed)&
 \tol{2.1500080(1)}{}{}&
 \tol{2.150008}{}{} (fixed)&
 \tol{2.15000825(12)}{}{} \\

 $T_\mathrm{C_1}-2456185 [\mathrm{BJD}_{\mathrm{UTC}}] $ &
\tol{0.602987(110)}{}{}&
\tol{0.60294(918)}{}{}\tablefootmark{a}&
\tol{0.60337(19)}{}{}\tablefootmark{a}&
	\tol{0.60247(19)}{}{}\tablefootmark{a}\\
 
$e$ 						&
\tol{0.0}{}{}  (fixed) &
\tol{0.0}{}{}  (fixed)&
\tol{0.0}{}{}  (fixed)&
\tol{0.0}{}{}  (fixed)\\

$u_1$ 					&
\tol{0.1993}{+0.0420}{-0.0430} &
$\tol{I~\mathrm{ band}=0.2045}{}{}$&
\tol{0.279}{+0.070}{-0.074}&
$\tol{R \mathrm{~band}= 0.28}{}{}$
\\

$u_2$ 					&
\tol{0.2862}{+0.0928}{-0.0974} &
$\tol{I \mathrm{~band}=0.3593}{}{}$&
\tol{0.254}{+0.123}{-0.122}&
$\tol{R \mathrm{~band}= 0.35}{}{}$
\\

$T_\mathrm{C_2}-2456200 [\mathrm{BJD}_{\mathrm{UTC}}]$\tablefootmark{a}&
\tol{0.65304(11)}{}{}&
\tol{0.65299(90)}{}{}&
\tol{0.65343(19)}{}{}&
\tol{0.65253(19)}{}{}
	 \\
	  \hline
\end{tabular}
\caption{Best fit planet system parameters from $\chi^2$ optimization and confidence intervals from MCMC sampling of the posterior parameter distributions for the Run 1 white light curve of HAT-P-32Ab. The red noise factor used to inflate the error bars of our results is $\beta=1.345$. For comparison the planet parameters obtained from three recent works \citep{2011ApJ...742...59H} ($e\equiv0$), \citep{2013MNRAS.436.2974G} and \citep{2014MNRAS.441..304S} are listed.\newline
{\tablefoottext{a}{Value derived from the ephemeris information given in respective papers.}}}
  \label{tab:wlresult}
\end{center}
\end{table*} 
The results for the white light curve fit to the data from Run 1 are listed in Table \ref{tab:wlresult} together with the literature parameters of recent high precision studies of HAT-P-32Ab. The error bars of these results were inflated with the red noise factor of  $\beta=1.345$ derived in Sect. \ref{sec:wlerror}. We find our results for the planet parameters to be consistent with the literature values. The difference in radius ratio compared to the results of \citet{2011ApJ...742...59H} and \citet{2014MNRAS.441..304S} might be explained by the different wavelength region probed in this study but is more likely caused by the transit depth dilution from additional flux of the M-dwarf companion HAT-P-32B, which was not accounted for in these two studies.\newline
A rough dilution correction of the given radius ratios using
\begin{equation}
\left(R_p/R_{\star}\right)_{\mathrm{corrected}}=\left(R_p/R_{\star}\right)_{\mathrm{uncorrected}}\cdot\sqrt{1+c_{\mathrm{filter}}}
	\label{eq:roughdi}
\end{equation}
with the dilution factor values $c_i'=0.006 \pm 0.002$, $c_z'=0.012 \pm 0.004$, $c_g'<0.0018$ provided by \citet{2014ApJ...796..115Z}  in similar broadband filters as the ones used in these two studies, yields comparable radius ratios to the one derived by us and \citet{2013MNRAS.436.2974G}.

\subsection{Transit timing comparison}\label{sec:ttv}
In order to compare our best fit transit time, $T_{\mathrm{C}_1,2456185}$ with literature predictions we use the ephemeris given in each paper to calculate the expected transit time and its uncertainty using Eq. \ref{eq:ephem}. 
\begin{equation}
T_{\mathrm{C}}=T_{0}+E\cdot P
	\label{eq:ephem}
\end{equation}
Where $T_{0}$ is a reference transit time, $E$ the epoch and $P$ the orbital period given in the respective paper. We find that our result for the transit time best agrees with the prediction given by the discovery paper \citep{2011ApJ...742...59H}. Using their ephemeris information and our measured transit time to refine the period we yield $P_{\mathrm{new}}=2.15000806(24)$ days.\newline
We also calculated the predicted transit time for Run 2 $T_{\mathrm{C}_2,2456200}$, which we could not reliably measure from a white light curve transit fit. To calculate the prediction for $T_{\mathrm{C}_2,2456200}$ from our Run 1 results we used Eq. \ref{eq:ephem} with $T_{0}=T_{\mathrm{C}_1,2456185}$, $E=7$ and $P=P_{\mathrm{new}}$. 
The results are given in Table \ref{tab:wlresult}.

\subsection{Comparison with theoretical limb darkening}\label{sec:limbd}
As the limb darkening coefficients are wavelength-dependent and cannot be well compared with literature results obtained in different filters, we compared them to theoretical values. To derive these theoretical coefficients we calculated the wavelength-dependent theoretical limb darkening profiles for a star with the basic stellar properties of HAT-P-32A ([Fe/H] $=-0.04\pm0.08$, $\log g$ (cgs) $=4.33 \pm 0.01$, \citet{2011ApJ...742...59H}, and $T_{\mathrm{eff}} =6269\pm64$ K, \citet{2014ApJ...796..115Z}) by interpolation from the PHOENIX specific intensity spectra library by \citet{2013A&A...553A...6H}. We then weighted each wavelength with its actual contribution to the measured stellar flux during observation, taking into account the instrument response function and the telluric absorption and summed all contributing limb darkening profiles to derive the theoretical white light profile. We then renormalized the model information so that $\mu=0$ actually corresponds to the outer edge of the star i.e. the region where the mean optical depth corresponds to unity.  This is not the case for the raw model data due to the spherical symmetry assumption used in the PHOENIX code. We repeated this process varying the adopted stellar parameters for HAT-P-32A within their reported errors to explore the uncertainties of the intensity profile.  Finally the theoretical limb darkening coefficients and their errors were derived by fitting the intensity profiles with a quadratic limb darkening law.
We found that our best fit limb darkening coefficients were lower than the values predicted by the PHOENIX stellar models ($u_{1,\mathrm{ theory}}=0.340\pm0.056$ and $u_{2,\mathrm{ theory}}=0.245\pm0.073$). This difference might be caused by insufficiencies of the PHOENIX models, errors in the assumed stellar parameters or undetected systematic noise in the white light curve of Run 1.
\subsection{Consequences for the retrieval of the planet's transmission spectrum}
Despite the good agreement of our Run 1 white light curve results with the literature data, the lack of extensive out-of-transit data lead us to deem the data set unsuitable for a transmission spectroscopy study where a reliable measurement of very small changes in the transit depth is essential.\newline
Further, we considered the results obtained for the white light curve fit of the Run 2 as unreliable since a clear and dominant systematic noise signal in the data remained uncorrected (see Sect. \ref{sec:rednoise}). We will further investigate this noise signal in Sect. \ref{sec:systematics} and motivate a correction for the narrow band channel data of Run 2 which then can be used to derive a transmission spectrum of HAT-P-32Ab.

\section{GTC/OSIRIS instrument specific systematic noise affecting the data of Run 2}\label{sec:systematics}
\begin{figure}
  \resizebox{\hsize}{!}{\includegraphics{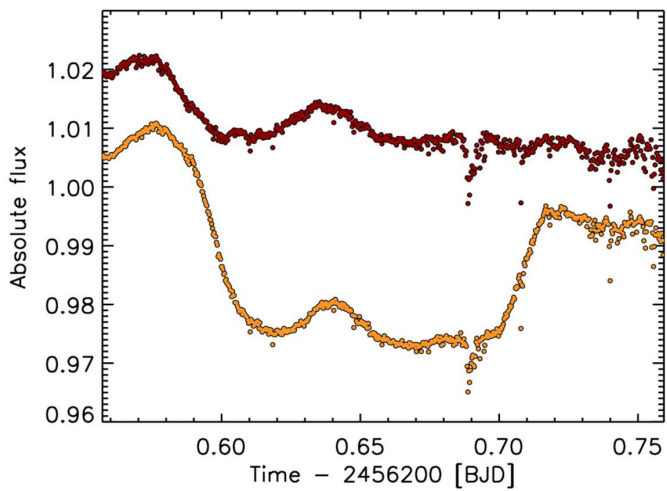}}	
  	\caption{Raw white light curves of HAT-P-32Ab (yellow) and the reference star Ref2 (red) from Run 2. A color version of this plot is available in the online version.}
	\label{fig:whitelight_h322_raw_curves}
\end{figure}
\begin{figure}
  \resizebox{\hsize}{!}{\includegraphics{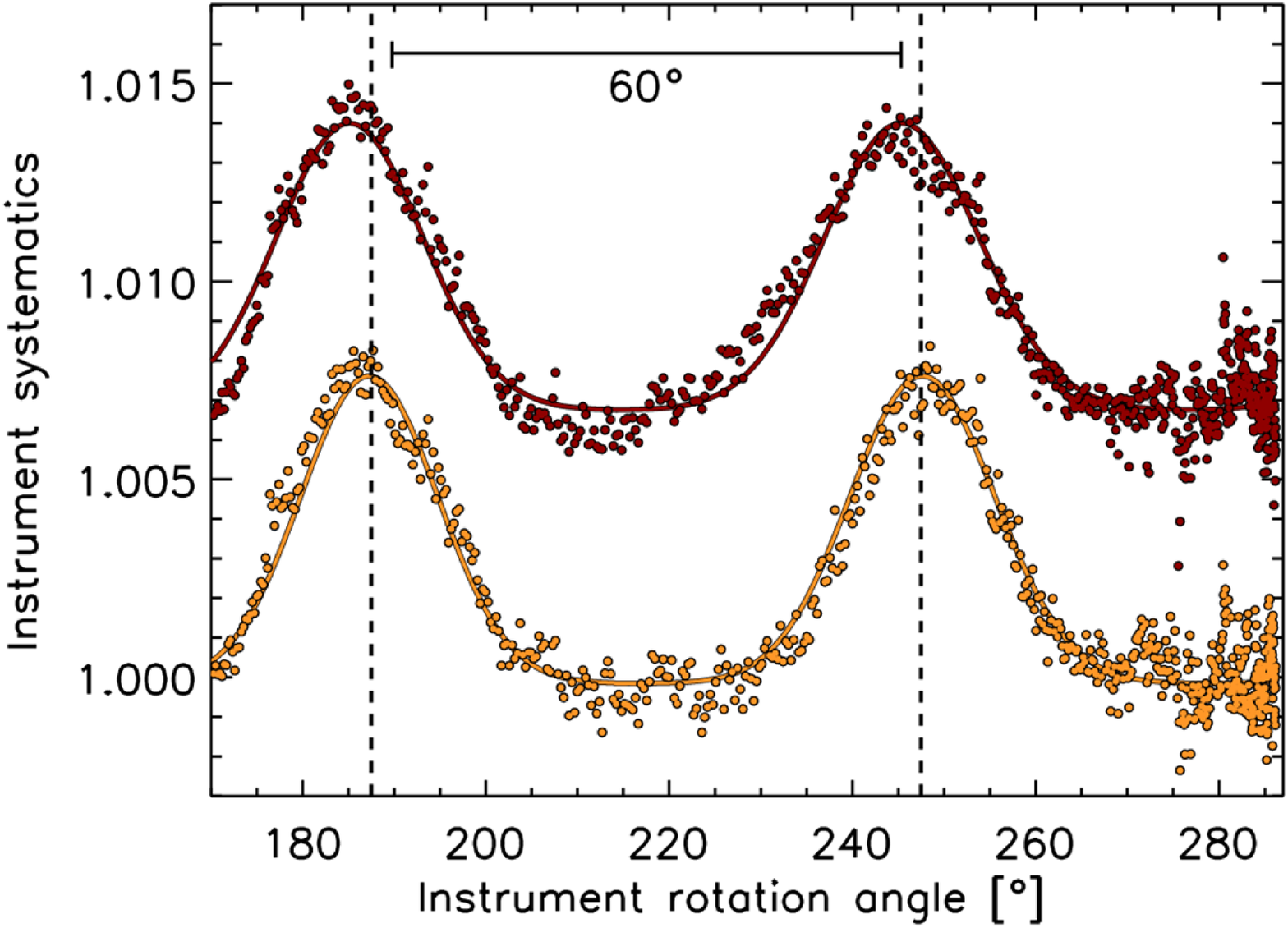}}	
  	\caption{Raw white light curves of HAT-P-32Ab (yellow circles) and Ref2 (red circles) from Run 2 plotted over the instrument rotation angle for better visualization of the systematic noise signal consisting of two `bumps'. Both curves were corrected by an air mass term and the curve of HAT-P-32A was divided by a theoretical transit. For both raw light curves a rough theoretical approximation of the data consisting of a sequence of two Gaussian functions is plotted over the data (yellow and red lines).}
	\label{fig:whitelight_h322_raw_curves_vs_rot}
\end{figure}
\begin{figure}
  \resizebox{\hsize}{!}{\includegraphics{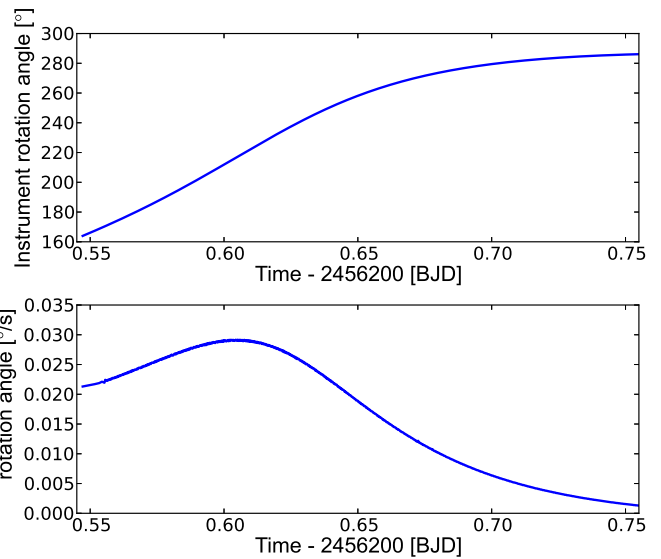}}	
  	\caption{\textit{Top panel:} Progression of the instrument rotation angle during the second observing night (Run 2). \textit{Bottom panel:} Speed of the changing instrument rotation angle during the second observing night (Run 2). The speed of the instrument rotation slows down in the second half of the observing run. }
	\label{fig:rotan_run2}
\end{figure}
\begin{figure}
  \resizebox{\hsize}{!}{\includegraphics{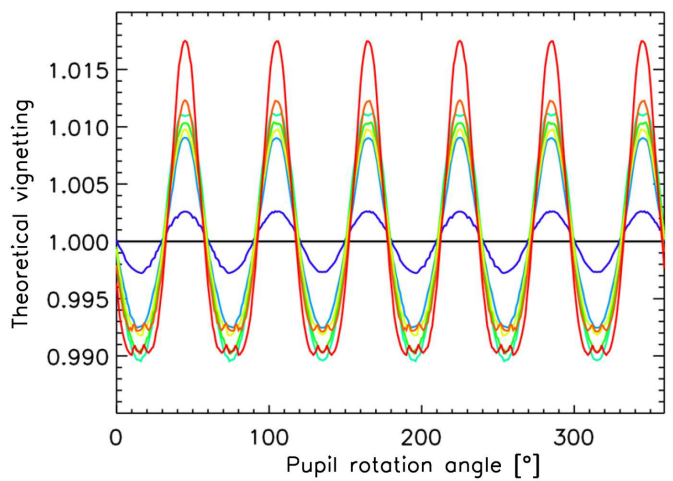}}	
  	\caption{Theoretical vignetting behavior as a function of effective pupil rotation angle derived from a simple toy model. The different colors represent different distances of the theoretically observed source from the telescope optical center (blue representing the closest distance and red the largest distance). A color version of this plot is available in the online version.}
	\label{fig:theo_vignetting}
\end{figure}
In Sect. \ref{sec:wlfit} we determined that the systematics affecting the data of Run 2 cannot be sufficiently corrected by solely using a similarly simple de-correlation function of auxiliary parameters as was sufficient for the data of Run 1. The largest non-astrophysical systematic present in the white light curve of Run 2 is the large distortion of the first half of the light curve. 
It is present in all narrow band channel light curves (see Fig. \ref{fig:raw_and_corrected_curves_blue}, left panel). When studying the raw white light curves of Run 2 we found that the distortion in the relative light curve coincides with a sinusoidal-like feature found in both the raw light curve of HAT-P-32A and the reference star (see Fig. \ref{fig:whitelight_h322_raw_curves}). This particular and slightly curious feature in the first half of the observation does not appear to be caused by telluric atmosphere variations. A closer investigation of the two `bumps' showed that while they are present in the data of both stars, they are slightly shifted in phase (and possibly have a different amplitude). As a consequence they do not cancel out when the ratio of both curves is taken to correct any telluric effects. We found that the `bumps' re-appear with a frequency of roughly $60\degr$ of the instrument rotation angle. To better visualize our  findings we divided the raw light curve of HAT-P-32A by a theoretical transit light curve using our best fit transit parameters from Run 1, cleaned both raw light curves from an air mass trend and plotted them over the instrument rotation angle (see Fig. \ref{fig:whitelight_h322_raw_curves_vs_rot}). We fitted all four `bumps' with a Gaussian function to derive an estimate for the values of rotation angle at which the peak flux occurs. Since the instrument rotation angle changed during the first half of the observing run and then stayed nearly constant (see Fig. \ref{fig:rotan_run2}) this lead to the `bumps' only manifesting in the first half of the run. The $60\degr$ symmetry of the `bump'-feature prompted us to suspect these flux amplitude variations to be caused by vignetting in pupil space. A vignetting of the pupil would reduce the overall amount of captured light that reaches the detector.  Due to the hexagonal shape of the GTC primary mirror the pupil also is not invariant under rotation but exhibits a  $60\degr$ rotational symmetry.  As a consequence, the amount of vignetted pupil area would not only depend on the distance of the projected pupil but also on its effective rotation angle to the source of the vignetting. Vignetting could for example occur at the secondary and/or tertiary mirror for off-axis rays. It is indicated in the GTC conceptual design document \citep[][Fig. 4.4]{GTCCD} that this should only have a significantly impact on targets located at separations larger than $\approx4\arcmin$ from the telescope optical axis (which includes Ref2 but not HAT-P-32A in Run 2, see Fig. \ref{fig:setup}). Potential vignetting caused by other objects located in the optical path in pupil space might, however, also affect targets located closer to the telescope pointing.\newline
We constructed a simple toy model, to simulate the expected footprint of pupil vignetting at a hypothetical  source. The effect predicted by this toy model exhibited a $60\degr$-periodicity with rotation angle very similar to the signal found in our data of Run 2 (see Fig. \ref{fig:theo_vignetting}). The toy model further showed a dependency of the signal on the initial target position relative to the telescope optical axis, which was regulating the phase, amplitude and precise shape of the effect. The true signal affecting the data could be more complex in case the source of vignetting is located off-center from the telescope optical axis (divergent from the toy model assumption) or if the signal is a superposition of several vignetting signals.\newline
The difference in systematic noise signal amplitude between the two stars of Run 2, which are separated by more than $3\arcmin$ ($\equiv$ 1/3 of the OSIRIS field of view) appears to be small, suggesting that for any given observing set up no signals with significantly larger amplitudes (i.e. larger than $1\%$) are to be expected. Systematic noise signals of amplitudes slightly lower than $1\%$ might be negligible for many other science cases. In our case, however, noise signals of this amplitude are more than two orders larger than the astrophysical signals we are aiming to detect and, therefore, need to be addressed. Since the amount of vignetting is only dependent on the projected pupil position and effective pupil rotation angle it should be wavelength-independent as long as the vignetting occurs before the light has passed any dispersing optical elements. In this case the produced systematic noise signal will affect all narrow band channel light curves and the white light curve in exactly the same way and should divide out if the ratio between any color light curve and the white light curve is taken.

\section{Analysis of the color light curves of Run 2}\label{sec:colorfit}
\begin{figure}
  \resizebox{\hsize}{!}{\includegraphics{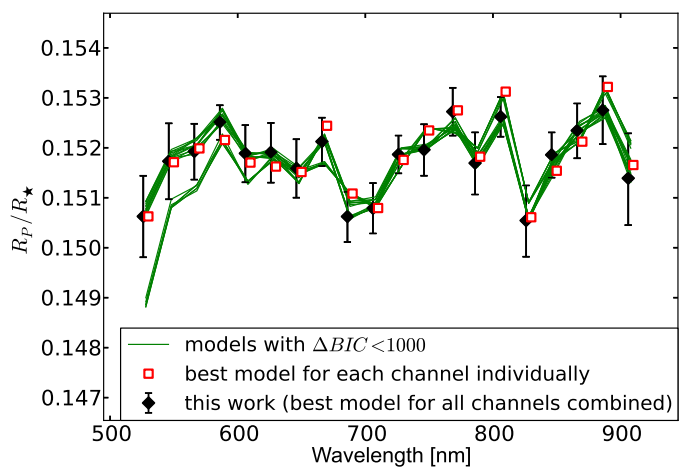}}	
  	\caption{Results for the wavelength-dependent radius-ratios obtained the different models listed in Table \ref{tab:bich322} (green lines). The final adapted results calculated with a homogeneous analysis are shown on top as black diamond (error bars are calculated as described in Sect. \ref{sec:errh22}). The results obtained when using the individually best suited model for each wavelength channel are plotted as red squares.}
	\label{fig:bicrprs}
\end{figure}

We proceeded by dividing all narrowband light curves of Run 2 by the white light curve. This served to cancel out all wavelength-independent systematic noise signals, i.e. signals which affected all curves in an identical manner. Such common-mode corrections are usual practice when dealing with spectrophotometric transit data \citep{2013MNRAS.436.2974G,2015MNRAS.446.2428S}. In accordance with expectation the correction appears to fully dispose of the light curve distortion we suspect to be caused by the wavelength-independent pupil vignetting (see Sect. \ref{sec:systematics}). All remaining (i.e. wavelength-dependent) noise signals in the narrow band light curves are found to be linked with auxiliary parameters of the observation, in a similar manner as it was the case for the white light curve of Run 1 (see Sect. \ref{sec:wlfit}). We fitted all differential light curves with a model of the form:
\begin{equation}
	\mathcal{M}_{\mathrm{diff},n}=\left(\mathcal{T}_n+c_n\right)/\left(\mathcal{T}_w+c_w\right)\cdot\mathcal{S}_n
	\label{eq:mdiff}
\end{equation}
where $\mathcal{T}_n$ is the analytical transit model for the narrow band light curve, $c_n$ the dilution factor (i.e. relative flux contribution of the stellar companion HAT-P-32B) in this particular narrow band, $\mathcal{T}_w$ the model for the white light transit and $c_w$ dilution factor in white light and $\mathcal{S}_n$ a systematic noise model for all remaining wavelength-dependent noise sources. The latter was of the same form as the systematic model used during the white light curve fit, i.e. a combination of polynomial functions of the auxiliary parameters detector position, air mass and FWHM.\newline 
Alternatively the original relative light curves can be fitted by a model of the form:
\begin{equation}
\mathcal{M}_n=\left(\mathcal{T}_n+c_n\right)\cdot \mathcal{S}_{\mathrm{CM}}\cdot\mathcal{S}_n
\end{equation}
where 
\begin{equation} \mathcal{S}_{\mathrm{CM}}=\frac{\mathcal{D}_w}{\left(\mathcal{T}_w+c_w\right)} \end{equation}
is the common mode systematic noise signal derived from the residuals of the white light curve data $\mathcal{D}_w$ of Run 2. This approach is mathematically identical to the one given in Eq. \ref{eq:mdiff}.\newline
In both cases all wavelength-independent transit model parameters (i.e. $a_p/R_{\star}$, $i$, $T_\mathrm{C_2}$, $P$, the white light planet-to-star radius ratio, the white light limb darkening coefficients and the dilution parameters $c_w$ and $c_n$), were kept fixed in this optimization. The values for $c_n$ were determined as explained in Appendix \ref{sec:sec}. We chose to fix the white light and wavelength-independent transit parameters to the values we derived from Run 1 in Sect. \ref{sec:wlfit}, as they represent a measurement taken with the same instrument and at the same wavelength interval. Not fixing the values but letting them vary in their uncertainty intervals did not change our final results for the relative wavelength-dependent change in radius ratio, but did only offset the exact value around which the narrow band radius ratios and limb darkening coefficients varied. The same held true when we fixed the wavelength-independent and white light transit parameters to the literature values provided by \citet{2013MNRAS.436.2974G}, \citet{2011ApJ...742...59H} or \citet{2014MNRAS.441..304S}. Between these tested configurations adopting our best fit Run 1 white light parameters did yield the lowest overall $\chi^2$. We, therefore, adopted the values resulting from these white light curves values as our final results. To allow an independent use of our derived transmission spectrum, we also provide the relative change in transit depth with respect to the white light curve transit depth, i.e. $\left(R_p/R_{\star}\right)^2-\left(R_p/R_{\star}\right)^2_\mathrm{white~light}$, together with our results for the absolute values of the radius-ratios in Table \ref{tab:tspec_results}. These differential transit depths are independent of the white light curve parameters chosen during the analysis and, therefore, free of any possible systematic errors caused by the uncertainties inherent to these parameters.\newline
We again tested all different combinations of systematics models. This time we allowed the highest polynomial order for the air mass and spatial and dispersion position to be 2 and for the FWHM to be 3. The $BIC$ was again calculated using Eq. \ref{eq:bic}. As we were aiming to determine the model which best explained all 20 channels simultaneously, the squared sum of all 20 channel's joined light curve residuals (i.e. channel \#1 - \#20, excluding the sub channel \#13b) were used for the calculation. The overall lowest $BIC$ was reached by using the model:
\begin{equation}
\begin{split}
	\mathcal{S}_n=~&n_0+x_1\cdot xpos+y_1\cdot ypos+	a_1\cdot airm\\
&	f_1\cdot fwhm+f_2\cdot fwhm^2
	\end{split}
\end{equation}
where $n_0$, $x_1$, $y_1$, $a_1$ and $f_{1,2}$ are wavelength-dependent parameters different for every channel.
An overview of the $BIC$ comparison can be found in Table \ref{tab:bich322} where all models with a $\Delta BIC < 1000$ are listed. 
\begin{table}
\begin{center}
\begin{tabular}{c c}
\hline\hline
Model &$\Delta {BIC}$\\ \hline
      xpos0 ypos1 airm1 fwhm2 &  114.82 \\
      xpos0 ypos1 airm1 fwhm3 &  211.20 \\
      xpos0 ypos2 airm1 fwhm2 &  282.08 \\
      xpos1 ypos1 airm1 fwhm2 &    \textbf{0.00} \\
      xpos1 ypos1 airm1 fwhm3 &   77.08 \\
      xpos1 ypos2 airm1 fwhm2 &  170.87 \\
      xpos1 ypos2 airm1 fwhm3 &  240.33 \\
      xpos2 ypos1 airm1 fwhm2 &  179.62 \\ 
      xpos2 ypos1 airm1 fwhm3 &  255.41 \\
      xpos2 ypos2 airm1 fwhm2 &  350.57 \\
      xpos2 ypos2 airm1 fwhm3 &  418.44 \\ \hline
\end{tabular}
\caption{Model comparison for the systematic noise in the narrow band channel curves of Run 2 listing the respective $\Delta  {BIC}= {BIC}- {BIC}_{ \mathrm{min}}$ values.}  \label{tab:bich322}
\end{center}
\end{table}
When investigating the individual channels separately we found that most of the color light curves were satisfied with a less complex model, without a spatial position ($xpos$) dependent term. But the $\chi^2$ of the three bluest channels was significantly minimized by introducing this additional parameter, leading to an overall favored $BIC$ without affecting the final results for the other channels.\newline
The results from the different models listed in Table \ref{tab:bich322} are plotted in Fig. \ref{fig:bicrprs} as green lines. The values we adapted as our final results are shown as black diamonds. It can be seen that at the blue end of the spectrum a few model give notably different results. The models yielding such deviant results are, however, not the ones favored when determining the best model for each channel individually. The resulting radius-ratios for the individually determined best models (red squares) are in good agreement with the adopted results derived from the homogeneous analysis.
\subsection{Errors}\label{sec:errh22}
\begin{figure*}
  \resizebox{\hsize}{!}{\includegraphics{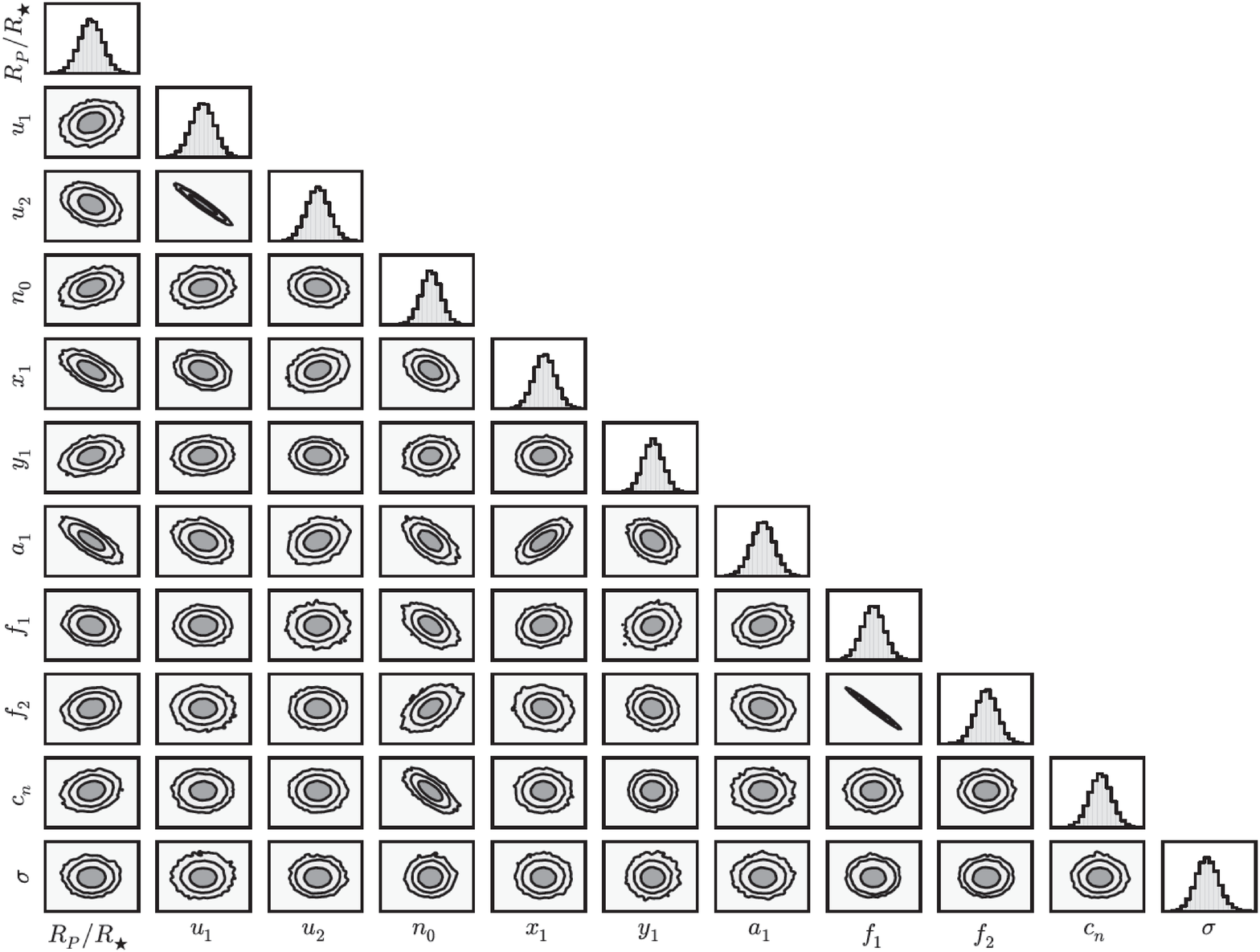}}	
  	\caption{Example correlation plots for the posterior parameter distributions for the fit of narrow band channel \#9 (678 nm -  698 nm). The lines indicate the areas encompassing $1\sigma$, $2\sigma$ and $3\sigma$ of the whole distribution, respectively.}
	\label{fig:confidence_intervals_h322_bin8}
\end{figure*}

\begin{figure}
\resizebox{\hsize}{!}{\includegraphics{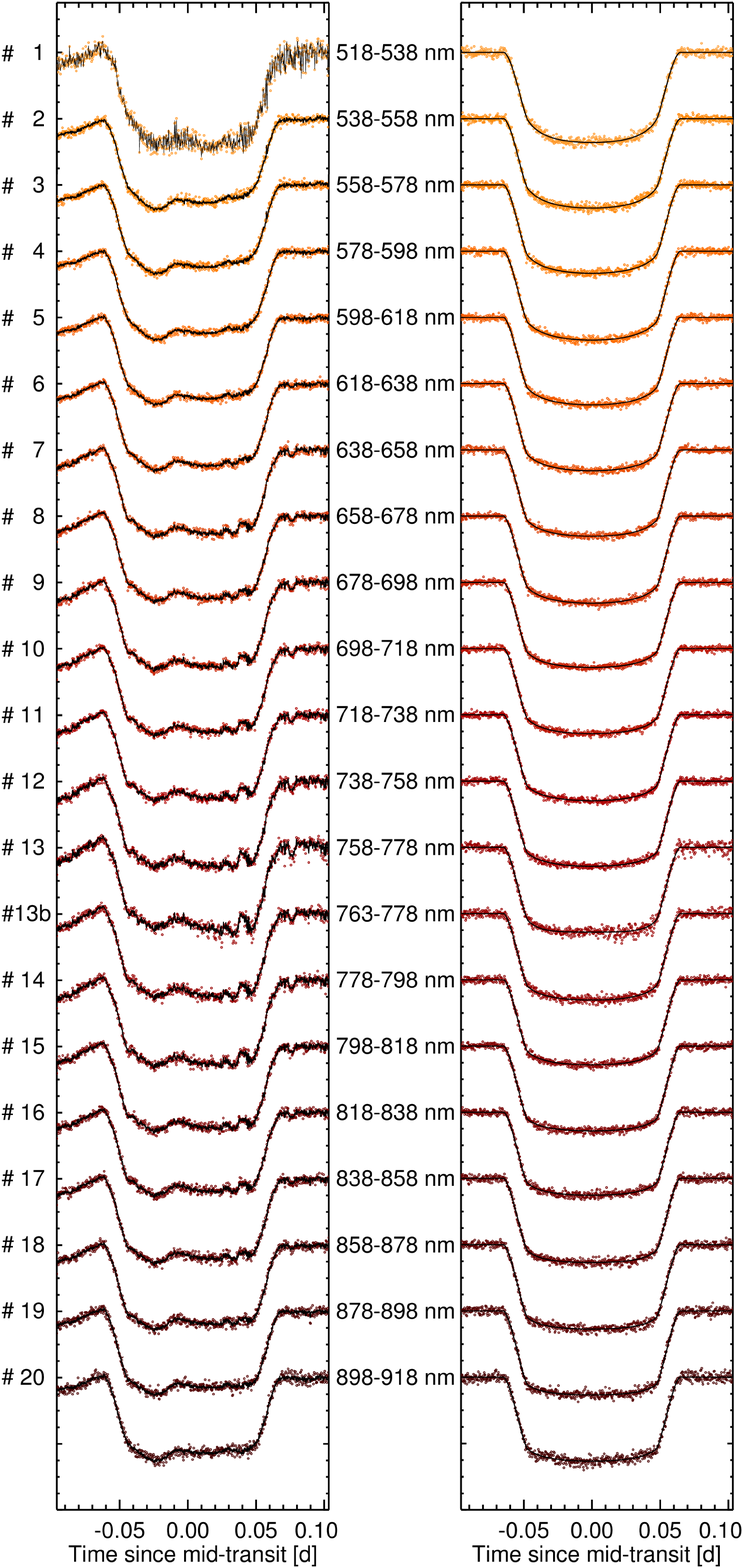}}
  	\caption{All narrow band channel light curves before and after correction from the systematic noise signals plotted together with the best fit model (black line).}
	\label{fig:raw_and_corrected_curves_blue}
\end{figure}

We repeated the exploration of the posterior parameter distributions with MCMC as described for the white light curve analysis in Sect. \ref{sec:wlerror}, now letting the values for the wavelength-dependent dilution factors $c_n$ vary within their uncertainty intervals using Gaussian priors. This time we used 300 walkers and let them run for 10\,000 accepted steps. We discarded all points of the burn-in period, which was over after 1000 steps. The typical thinning factors for the chains were $\approx 82$, leaving us with a final distribution of an average of 33\,300 points for each wavelength channel light curve fit. An example of the resulting posterior parameter distributions is show for channel \#9 in Fig. \ref{fig:confidence_intervals_h322_bin8}. We then calculated the red noise factor for every narrow band channel light curve as described in Sect. \ref{sec:rednoise} and inflated the MCMC derived error bars accordingly. The resulting values for $\beta$ lie between $1.022$ and $1.950$ and the resulting uncertainties of the wavelength-dependent radius ratio after multiplication with $\beta$ lie between $337$ and $972$ ppm. The light curves before and after the full systematic noise correction are shown in Fig.\ref{fig:raw_and_corrected_curves_blue}. The resulting radius ratios and limb darkening coefficients as well as $\beta$ factors are given in Table \ref{tab:tspec_results}.
\def\tol#1#2#3{\hbox{\rule{0pt}{15pt}${#1}^{{#2}}_{{#3}}$}}
 \begin{sidewaystable*}
\begin{center}
\begin{tabular}{cccccccc}
\hline\hline
Channel& $\lambda$ [nm]&
 $R_p/R_{\star}$& $\Delta$ Transit depth $\left[\%\right]$&$u_1$&$u_2$&Dilution factor $c_n$ $\left[\%\right]$&Red noise factor $\beta$\\ \hline
 \#1&$         518 -          538$&\tol{    0.150627}{+    0.000812}{   -0.000818}&\tol{    -0.03055}{+     0.02453}{    -0.02458}&\tol{       0.272}{+       0.054}{      -0.054}&\tol{       0.461}{+       0.089}{      -0.089}&$      0.1574\pm      0.0522$&$       1.862$\\ 
\#2&$         538 -          558$&\tol{    0.151731}{+    0.000759}{   -0.000761}&\tol{     ~~\,0.00283}{+     0.02309}{    -0.02304}&\tol{       0.280}{+       0.050}{      -0.051}&\tol{       0.331}{+       0.083}{      -0.082}&$      0.1689\pm      0.0565$&$       1.950$\\ 
\#3&$         558 -          578$&\tol{    0.151927}{+    0.000553}{   -0.000564}&\tol{     ~~\,0.00878}{+     0.01683}{    -0.01711}&\tol{       0.256}{+       0.038}{      -0.039}&\tol{       0.305}{+       0.063}{      -0.062}&$      0.1796\pm      0.0541$&$       1.612$\\ 
\#4&$         578 -          598$&\tol{    0.152515}{+    0.000337}{   -0.000355}&\tol{     ~~\,0.02669}{+     0.01029}{    -0.01082}&\tol{       0.286}{+       0.023}{      -0.023}&\tol{       0.242}{+       0.038}{      -0.038}&$      0.1710\pm      0.0553$&$       1.022$\\ 
\#5&$         598 -          618$&\tol{    0.151889}{+    0.000565}{   -0.000578}&\tol{     ~~\,0.00763}{+     0.01720}{    -0.01752}&\tol{       0.248}{+       0.039}{      -0.040}&\tol{       0.270}{+       0.065}{      -0.065}&$      0.1944\pm      0.0554$&$       1.761$\\ 
\#6&$         618 -          638$&\tol{    0.151910}{+    0.000588}{   -0.000607}&\tol{     ~~\,0.00827}{+     0.01790}{    -0.01840}&\tol{       0.251}{+       0.041}{      -0.041}&\tol{       0.242}{+       0.067}{      -0.067}&$      0.1855\pm      0.0525$&$       1.950$\\ 
\#7&$         638 -          658$&\tol{    0.151590}{+    0.000582}{   -0.000588}&\tol{    -0.00144}{+     0.01768}{    -0.01779}&\tol{       0.222}{+       0.040}{      -0.041}&\tol{       0.275}{+       0.066}{      -0.065}&$      0.3335\pm      0.0601$&$       1.908$\\ 
\#8&$         658 -          678$&\tol{    0.152125}{+    0.000478}{   -0.000489}&\tol{     ~~\,0.01481}{+     0.01457}{    -0.01485}&\tol{       0.239}{+       0.033}{      -0.033}&\tol{       0.249}{+       0.053}{      -0.054}&$      0.2665\pm      0.0587$&$       1.575$\\ 
\#9&$         678 -          698$&\tol{    0.150625}{+    0.000493}{   -0.000511}&\tol{    -0.03061}{+     0.01488}{    -0.01537}&\tol{       0.207}{+       0.034}{      -0.036}&\tol{       0.290}{+       0.057}{      -0.056}&$      0.2700\pm      0.0606$&$       1.702$\\ 
\#10&$         698 -          718$&\tol{    0.150789}{+    0.000504}{   -0.000504}&\tol{    -0.02566}{+     0.01522}{    -0.01517}&\tol{       0.213}{+       0.036}{      -0.036}&\tol{       0.267}{+       0.058}{      -0.058}&$      0.3621\pm      0.0583$&$       1.738$\\ 
\#11&$         718 -          738$&\tol{    0.151867}{+    0.000378}{   -0.000378}&\tol{     ~~\,0.00696}{+     0.01150}{    -0.01147}&\tol{       0.214}{+       0.026}{      -0.026}&\tol{       0.242}{+       0.043}{      -0.043}&$      0.5098\pm      0.0646$&$       1.285$\\ 
\#12&$         738 -          758$&\tol{    0.151961}{+    0.000510}{   -0.000520}&\tol{     ~~\,0.00982}{+     0.01553}{    -0.01578}&\tol{       0.204}{+       0.036}{      -0.036}&\tol{       0.210}{+       0.059}{      -0.057}&$      0.7963\pm      0.0613$&$       1.626$\\
(\#13&$        758 -          778$&\tol{    0.152514}{+    0.000972}{   -0.000786}&\tol{     ~~\,0.02666}{+     0.02974}{    -0.02391}&\tol{       0.005}{+       0.072}{      -0.011}&\tol{       0.538}{+       0.022}{      -0.116}&$      0.6573\pm      0.0588$&$       1.731$)\\  
\#13b&$         763 -          778$&\tol{    0.152724}{+    0.000476}{   -0.000480}&\tol{     ~~\,0.03307}{+     0.01456}{    -0.01464}&\tol{       0.171}{+       0.034}{      -0.034}&\tol{       0.297}{+       0.055}{      -0.054}&$      0.6211\pm      0.0568$&$       1.191$\\ 
\#14&$         778 -          798$&\tol{    0.151694}{+    0.000614}{   -0.000629}&\tol{     ~~\,0.00171}{+     0.01867}{    -0.01904}&\tol{       0.182}{+       0.044}{      -0.045}&\tol{       0.262}{+       0.073}{      -0.070}&$      0.7601\pm      0.0644$&$       1.886$\\ 
\#15&$         798 -          818$&\tol{    0.152622}{+    0.000393}{   -0.000400}&\tol{     ~~\,0.02995}{+     0.01201}{    -0.01219}&\tol{       0.159}{+       0.028}{      -0.029}&\tol{       0.230}{+       0.047}{      -0.046}&$      1.0250\pm      0.0684$&$       1.212$\\ 
\#16&$         818 -          838$&\tol{    0.150545}{+    0.000701}{   -0.000725}&\tol{    -0.03302}{+     0.02116}{    -0.02178}&\tol{       0.149}{+       0.051}{      -0.052}&\tol{       0.282}{+       0.084}{      -0.082}&$      1.0954\pm      0.0670$&$       1.994$\\ 
\#17&$         838 -          858$&\tol{    0.151860}{+    0.000449}{   -0.000457}&\tol{     ~~\,0.00675}{+     0.01366}{    -0.01386}&\tol{       0.127}{+       0.032}{      -0.033}&\tol{       0.304}{+       0.052}{      -0.051}&$      1.0714\pm      0.0688$&$       1.230$\\ 
\#18&$         858 -          878$&\tol{    0.152345}{+    0.000542}{   -0.000555}&\tol{     ~~\,0.02150}{+     0.01654}{    -0.01688}&\tol{       0.135}{+       0.039}{      -0.040}&\tol{       0.268}{+       0.065}{      -0.063}&$      1.1793\pm      0.0703$&$       1.380$\\ 
\#19&$         878 -          898$&\tol{    0.152751}{+    0.000679}{   -0.000677}&\tol{     ~~\,0.03389}{+     0.02079}{    -0.02064}&\tol{       0.130}{+       0.048}{      -0.050}&\tol{       0.225}{+       0.081}{      -0.079}&$      1.2985\pm      0.0643$&$       1.633$\\ 
\#20&$         898 -          918$&\tol{    0.151391}{+    0.000898}{   -0.000939}&\tol{    -0.00747}{+     0.02727}{    -0.02834}&\tol{       0.181}{+       0.065}{      -0.068}&\tol{       0.223}{+       0.109}{      -0.106}&$      1.4459\pm      0.0701$&$       1.789$\\ 
   \hline 
 \end{tabular}
\caption{Best fit results for the planet-to-star radius ratio and limb darkening coefficients $u_1$, and $u_2$ for the narrow band channel light curves of Run 2. For every channel the red noise factor $\beta$ that was used to inflate the MCMC derived error bars of the results is given. Also the respective dilution factor $c_n$ that was given as prior information for every channel is listed. Separately listed are changes in the transit depth with respect to the white light transit depth i.e. $\Delta$ transit depth $=\left(R_p/R_{\star}\right)^2-\left(R_p/R_{\star}\right)^2_\mathrm{white~light}$. These values are independent from the absolute value of the white light radius-ratio assumed during the analysis.}
 \label{tab:tspec_results}
 \end{center}
\end{sidewaystable*} 
 \section{Results and discussion of the transmission spectrum}\label{sec:results}
 \begin{figure}
\resizebox{\hsize}{!}{\includegraphics{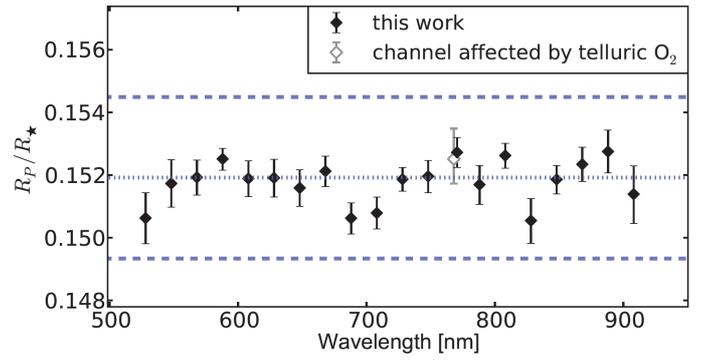}}
  	\caption{Results for the wavelength-dependent planet-to-star radius ratio of HAT-P-32Ab from Run 2. The blue dotted line indicates the mean radius ratio and the blue dashed lines indicate $\pm$ two atmospheric scale heights.}
	\label{fig:results_rprs}
\end{figure}
\begin{figure}
\resizebox{\hsize}{!}{\includegraphics{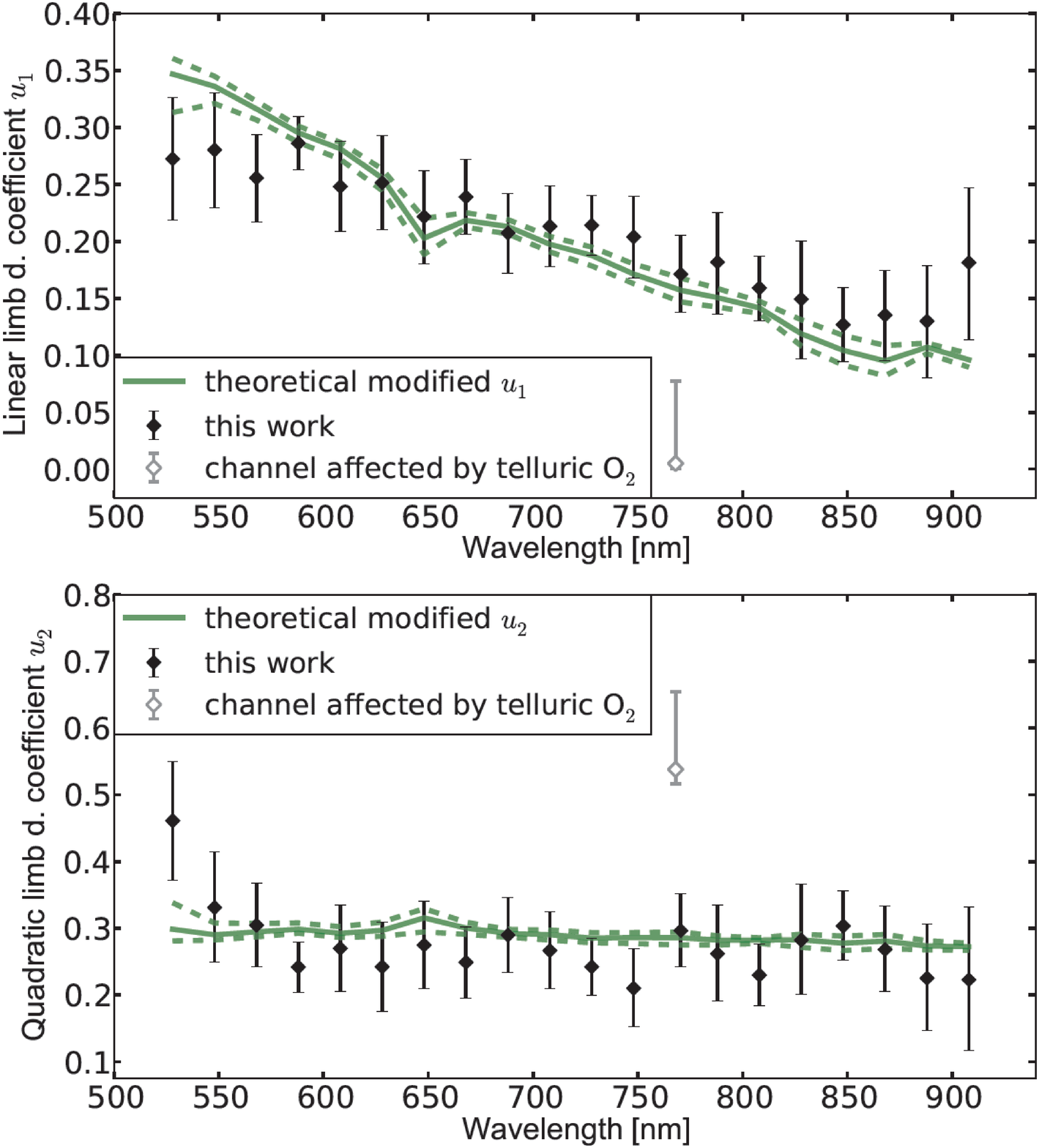}}
  	\caption{Progression of the results for the linear ($u_1$, \textit{top panel}) and quadratic ($u_2$, \textit{bottom panel}) limb darkening coefficients with wavelength from Run 2. Both are compared with a theoretical prediction derived from PHOENIX models (green line) and its error (green dashed line) taking into account the uncertainties in the stellar parameters of the host star HAT-P-32A.} 
	\label{fig:results_ld2}
\end{figure}
The result for the wavelength-dependent planet-to-star radius ratio of HAT-P-32Ab is shown in Fig \ref{fig:results_rprs}. We find it to show little variation in the probed region between $518$ and $918$ nm. The variations are significantly smaller than two planetary atmospheric scale heights ($H_p\approx1400$ km). This estimate for $H_P$ is derived with Eq. \ref{eq:scaleh} 
\begin{equation}
H_p=\frac{k_bT_p}{mg_p}\label{eq:scaleh}
\end{equation}
using the planet equilibrium temperature $T_p=2042$ K given by \citet{2014ApJ...796..115Z}, the planetary gravitational acceleration of $g_p=6.6069$ $\mathrm{m/s^2}$ from \citet{2011ApJ...742...59H} and the approximation of the mean molecular weight of the atmosphere $m$ as that of a solar abundance hydrogen helium mixture. In  Eq. \ref{eq:scaleh} $k_B$ stands for the Boltzmann constant. The best fit results for the limb darkening coefficients are shown in Fig. \ref{fig:results_ld2}. They vary smoothly with wavelength except for the coefficients for channel \#13, which encompasses the telluric oxygen bands at $\approx761$ nm. We compare the results for the limb darkening coefficients with theoretical predictions. Our approach to the narrow band channel analysis entailed the division by the white light curve. Therefore, effectively only relative changes of the transit depth and the stellar limb darkening to the white light transit depth and white light limb darkening were measured. Consequently, the absolute values of our resulting color dependent limb darkening values are affected by the assumed white light limb darkening values. Since we adopted the best fit white light curve parameters from Run 1, where the limb darkening coefficients are divergent from theoretical expectations, the resulting color dependent coefficients are by default divergent as well. In order to still independently compare them to the theoretical expectations we  calculate modified theoretical values which correspond to the theoretical predictions for the color dependent coefficients under assumption of the white light coefficients fixed to the best fit results of Run 1. These modified theoretical values were derived from PHOENIX specific intensity spectra as described in Sect. \ref{sec:limbd}, where instead of the whole white light wavelength region, now only the corresponding narrow band channels were summed to derive the limb darkening profile.  The resulting modified theoretical prediction is shown together with the actual measured limb darkening coefficients in Fig. \ref{fig:results_ld2}. It can be seen that the strong change in limb darkening in respect to the neighboring wavelength regions that we see for channel \#13 is not expected by theory. Since the results for the sub-channel \#13b do not show the same divergence from theory, we conclude that the results of channel \#13 are affected by uncorrected noise due to the strong telluric absorption bands and are un-reliable. Channel \#13 does not carry any significant information that is not also represented by channel \#13b. Therefore, we decide to exclude the results for the planet-to-star radius ratio of channel \#13 in the following comparison to literature data and planet atmosphere model predictions.
 
\begin{figure}
\resizebox{\hsize}{!}{\includegraphics{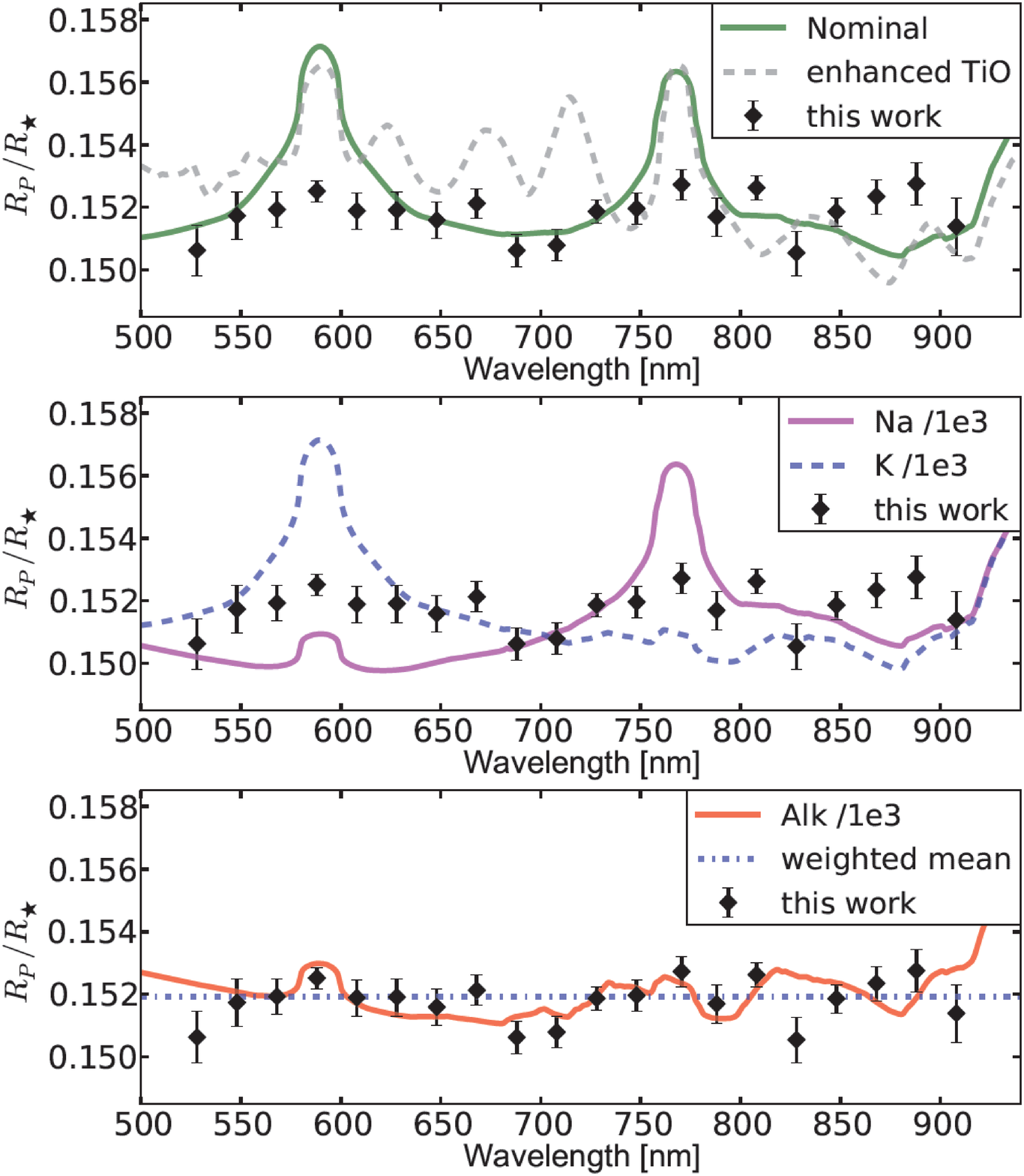}}
  	\caption{Results for the wavelength-dependent planet radius of HAT-P-32Ab from Run 2 (circles) compared with several different theoretical atmosphere models. \textit{Top panel:} Unmodified solar composition model at thermochemical equilibrium (green) and the same model with the TiO abundance enhanced in the upper layers of the atmosphere (grey dashed). \textit{Middle panel:} Models with selective alkali metal abundance depletion. In one model only sodium abundance was reduced  by a factor of 1000 (magenta) and in the other only potassium abundance was reduced  by a factor of 1000 (blue dashed). \textit{Bottom panel:} Model where both sodium and potassium abundance was reduced  by a factor of 1000 (red) and median average of the radius ratio values (blue dashed dotted). A color version of this plot is available in the online version.}  	
	\label{fig:compare_rprs_models}
\end{figure}

\begin{figure}
\resizebox{\hsize}{!}{\includegraphics{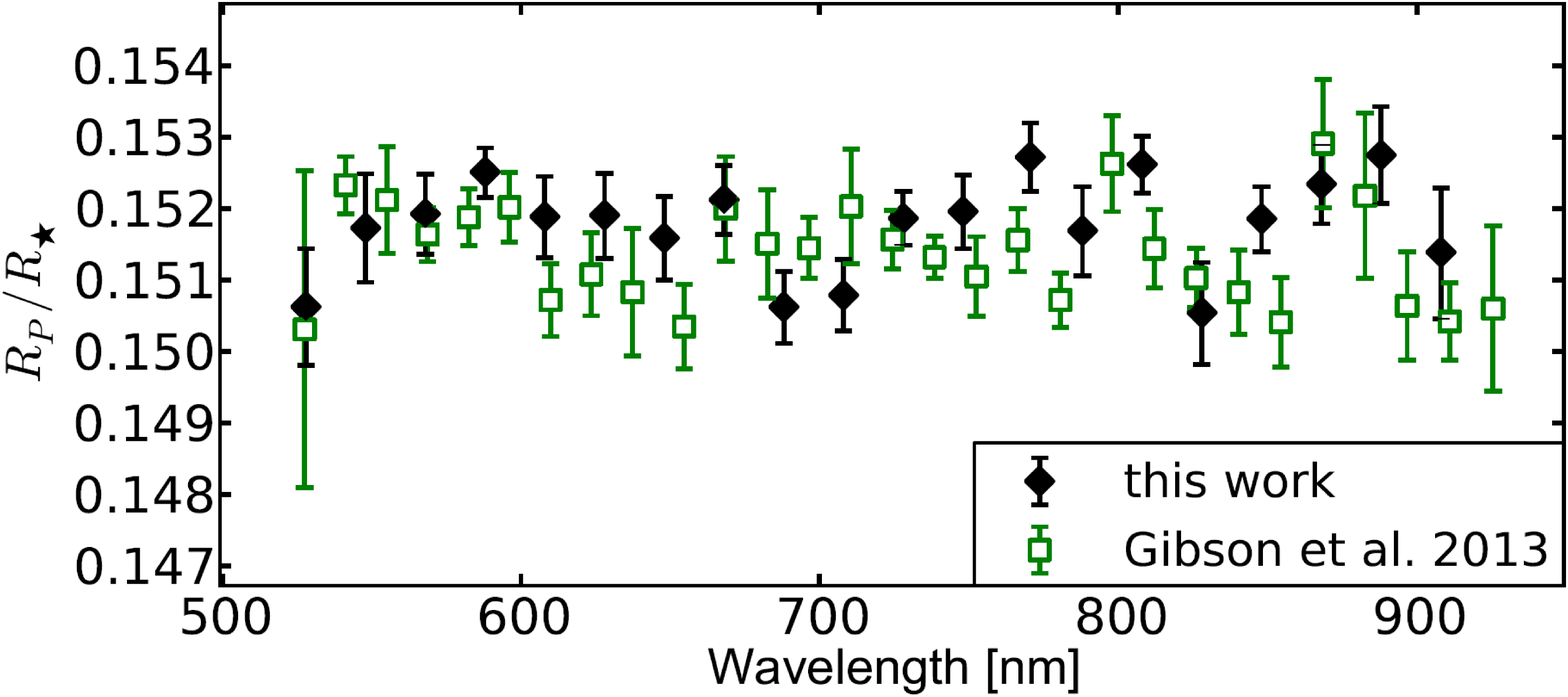}}
  	\caption{Results for the wavelength-dependent planet radius of HAT-P-32Ab from Run 2 compared with the results from \citet{2013MNRAS.436.2974G}.}
	\label{fig:compare_rprs_g13}
\end{figure}
\subsection{Comparison to theoretical models and literature data}
We compared the results to theoretical atmosphere models with various degrees of alkali metal abundance depletion (see Fig. \ref{fig:compare_rprs_models}). The atmospheric models were calculated with a line by line radiative code described in \citet{2005A&A...436..719I}, with updated opacities described in \citet{2012A&A...548A..73A} and \citet{2015ApJ...811...55M}. We assumed a clear atmosphere (no clouds) in thermochemical equilibrium with a solar abundance of elements for the nominal model. Furthermore, other models with various changes in TiO and alkali abundances were explored. The stellar heating was provided using a stellar synthetic spectrum\footnote{(available at ftp://ftp.stsci.edu/cdbs/grid/ck04models)} of a G-type star with an effective temperature of 6000~K from \citet{2003IAUS..210P.A20C}. We found no agreement between the measured transmission spectrum and the expected absorption features of the sodium and potassium resonance doublets (\ion{Na}{I} at 589.0 and 589.6 nm, \ion{K}{I} at 766.5 and 769.9 nm, with the doublets being unresolved at our resolution) predicted by the nominal model (see top panel of Fig. \ref{fig:compare_rprs_models}, green model). The radius ratio in channel \#4 encompassing the sodium doublet is lower by $\approx12.9\sigma$ than the predicted value and the radius ratio in channel \#13b encompassing the potassium doublet is lower by $\approx7.4\sigma$. Further, there is no indication of a detection of the predicted broad wings of the sodium feature in the channels neighboring channel \#4. Neither an enhancement of the titanium oxide (TiO) abundance in the probed upper layers of the atmosphere (see top panel of Fig. \ref{fig:compare_rprs_models}, grey dashed model) nor a depletion of either only sodium or only potassium (see middle panel of Fig. \ref{fig:compare_rprs_models}) improves the fit between the model and data. Reducing the abundance of both these alkali metals by a factor of 1000 yields a model which agrees fairly well with the data with a reduced  $\chi^2$ of $1.58$ (20 degrees of freedom). This is a similar but slightly worse match than the one to a straight line representative of a grey atmosphere (see bottom panel of Fig. \ref{fig:compare_rprs_models}) which results in a reduced $\chi^2 $ of $1.50$ (20 degrees of freedom). 
A grey atmosphere signal could be caused by a high altitude cloud layer masking the fingerprint of the atmosphere below. Alternatively, a significantly lower atmospheric scale height of the terminator region than the measured equilibrium temperature of the planet would suggest could be responsible. The amplitude of all expected atmospheric features would shrink accordingly and in an extreme case they would be hidden in the uncertainties of the measurement. To be in very good agreement with the data the scale height would have to be lower than the current prediction by a factor of $6.8$. Such a decrease in scale height could for example be caused by either a lower than predicted terminator temperature ($\approx300$ K) or an increase of the assumed mean molecular weight by that factor (i.e. $m\approx 16$ kg/kmole; for comparison the mean molecular weight of water vapor is $m_{\rm H_2O}=18.02$ kg/kmole). A more plausible explanation, however, would be that both of these two regulating factors are divergent from the values assumed in the current scale height calculation.\newline
Our results for the planet-to-star radius ratio are in good agreement with the study of \citet{2013MNRAS.436.2974G} who observed two transit events of HAT-P-32Ab with GMOS at Gemini North. They analyzed both data sets separately and then combined the final results for the planet-to-star radius ratio of both data sets. In Fig. \ref{fig:compare_rprs_g13} we compare our results with their combined results.

\subsection{Potential of GTC/OSIRIS as a tool for transmission spectroscopy}
We found the light curves to be heavily affected by systematics which can, however, be modeled. Only the channel containing the telluric oxygen absorption band  at $\approx761$ nm was too heavily affected by noise to be sufficiently corrected. The data showed correlation with the position of the stars on the chip (and in the slit), with air mass (which is expected due to the difference in color between the planet host star and the reference star)  and with seeing. The latter correlation suggests slit losses affecting the planet host star and reference star differently. This problem can be overcome by choosing a larger slit width. We adopted this for the following runs \citep[e.g.][]{2016A&A...585A.114P}. While the rotation dependent distortion cannot be overcome, it will affect the data less when the changes in the rotation angle are small. It will also have a smaller impact when the two stars are close to each other on the chip since vignetting then should be similar for both stars. In such cases, i.e. when no obvious distortions can be seen in the data, disregarding their possible hidden existence could lead to systematically erroneous transit parameters, but should have negligible impact on the derived transmission spectrum, as long as only relative variations in the radius ratio are considered.

\section{Conclusion}\label{sec:conclusion} 
We were able to derive a high precision transmission spectrum for the inflated hot Jupiter HAT-P-32Ab showing no prominent absorption features and, thus, supporting the results of the earlier work by \citet{2013MNRAS.436.2974G}. The study allowed us to detect and understand the low-level instrument systematics affecting GTC/OSIRIS and will help to improve future measurements. The independent confirmation of ground-based results from a different ground-based facility affected by different systematic noise signals is a step towards re-establishing faith in the reliability of (ground-based) transmission spectroscopy measurements. We have confidence in the potential of ground-based facilities and GTC/OSIRIS in particular as an excellent tool for larger surveys.  

\begin{acknowledgements}
This work is based on observations made with the Gran Telescopio Canarias 
(GTC), installed in the Spanish Observatorio del Roque de los Muchachos of the 
Instituto de Astrofisica de Canarias, in the island of La Palma. It is partly financed by the Spanish Ministry of Economics and Competitiveness through projects ESP2013-48391-C4-2-R and ESP2014-57495-C2-1-R. LN acknowledges support from the DFG Graduiertenkolleg 1351 \textit{Extrasolar Planets and their Host Stars}. FM acknowledges the support of the French Agence Nationale de la Recherche (ANR), under the program ANR-12-BS05-0012 Exo-atmos. Many of the plots shown in this paper were made using Matplotlib
 \citep{Hunter2007}.  We would like to note that during a late stage of this paper's referee process another study of HAT-P-32b by \citet{2016arXiv160309136M} was uploaded to ArXiv. This study further confirms the measurement of a flat transmission spectrum for this planet's atmosphere.
\end{acknowledgements}

\begin{appendix}
\section{Companion HAT-P-32B}\label{sec:sec}
In \citeyear{2013AJ....146....9A} \citeauthor{2013AJ....146....9A} discovered an optical companion to HAT-P-32A using adaptive optics (AO) with Aries, a near infrared diffraction limited imager and spectrograph (PI: Don McCarthy), which is fed by the Multiple Mirror Telescope's (MMT) AO beam. The optical companion was later confirmed to be bound to the HAT-P-32 system by \citet{2015ApJ...800..138N} from proper motion measurements and AO imaging. HAT-P-32B was further characterized as an M-dwarf by \citet{2014ApJ...796..115Z} and \citet{2015ApJ...800..138N} who used near infrared broad band AO imaging to constrain its stellar parameters.\newline
Due to the close proximity of the two stars HAT-P-32A and B the flux of the latter was contributing to our measurements of the former. In order to correctly include this effect in our models we needed to determine the wavelength-dependent flux ratio between the two stars as precisely as possible. In the following we describe how we extracted this information from our GTC/OSIRIS spectra. Subsequently, we use this data and additionally obtained near infrared observations to derive improved stellar parameters for HAT-P-32B.

\subsection{Optical spectrum - GTC/OSIRIS}
\begin{figure}
 \resizebox{\hsize}{!}{\includegraphics{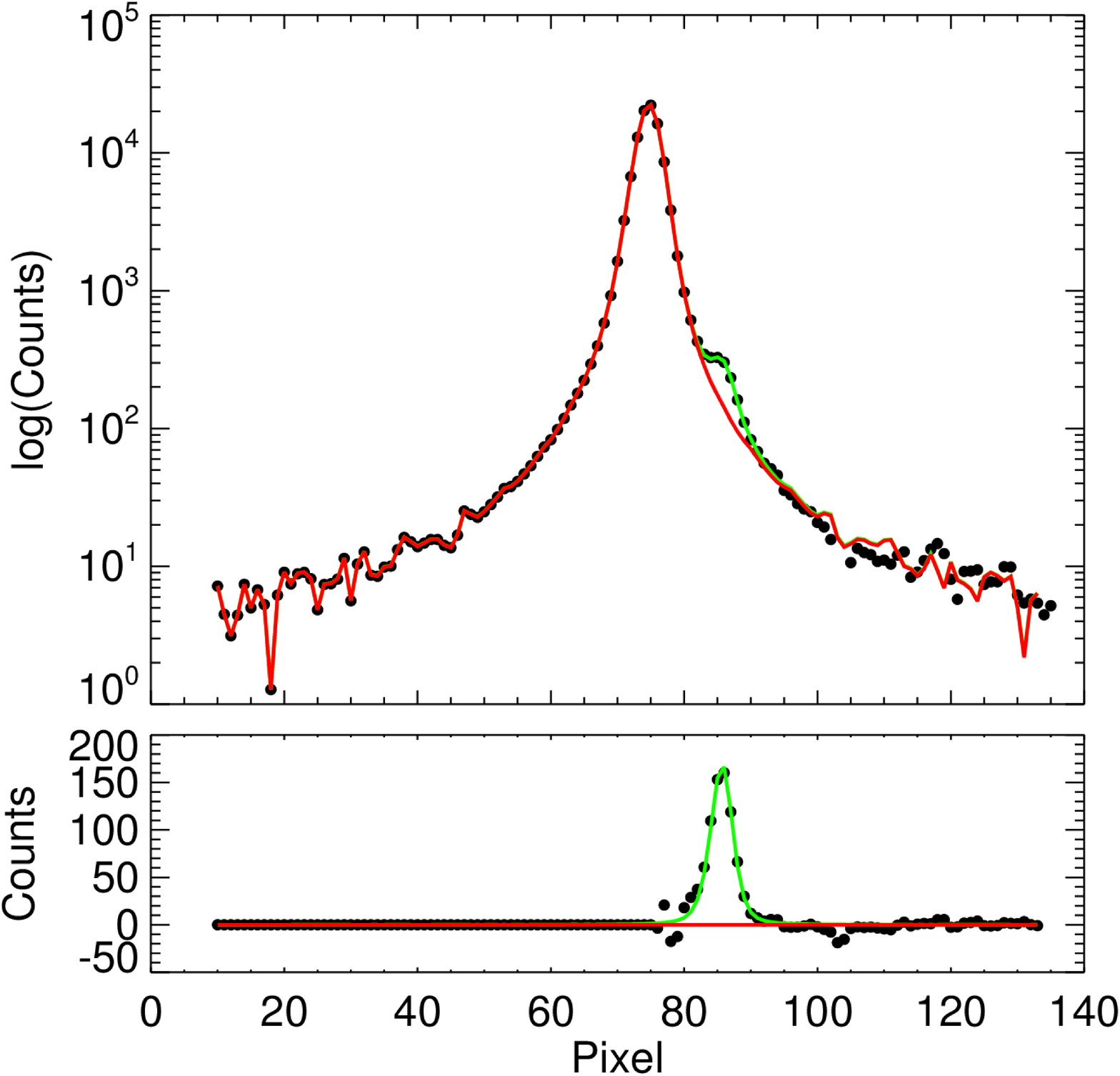}}
  	\caption{\textit{Top panel:}  Example for the combined stellar profiles of HAT-P-32A and HAT-P-32B plotted in semi logarithmic scale. In red the empirical model profile for HAT-P-32A is shown. The green line indicates where the combined profile diverges from HAT-P-32A's profile. \textit{Bottom panel:} Same data as in top panel after subtraction of the empirical model profile of HAT-P-32A. The green line now represents the model profile of HAT-P-32B. A color version of this plot is available in the online version.}
\label{fig:profile}
\end{figure}

Since HAT-P-32B was undiscovered prior to our observations the observing set up was not optimized to maximize the projected distance of the two spectra of HAT-P-32A and B on the chip. We still were able to detect HAT-P-32B as a separated object in Run 2 as a deformation of the spatial profile of HAT-P-32A.\newline 
Using the out-of-transit data available in Run 2 we determined the flux peak of the HAT-P-32A spectrum in every frame at every wavelength and then added frames in 10-frame time-bins resulting in 25 images with increased signal-to-noise. We then fitted a double-peak model based on an empirical profile function to the spatial profiles of HAT-P-32A and B for every wavelength cut though the spectrum for all 25 images. The used empirical profile was based on a Moffat function and assumed symmetry of the stellar profile. It is described in detail in Sect. \ref{sec:emprofile}. A sample fit to the spatial double profile can be found in Fig. \ref{fig:profile}.

\subsubsection{Empirical profile}\label{sec:emprofile}
We acquired a first approximation of the peak position and amplitude of the two stellar profiles by fitting each of them with a Moffat function enforcing identical values for the half widths and for the Moffat indices of both profiles. The result showed that the wings of the actual stellar profiles were underestimated by the Moffat approximation. Consequently, we moved on to an empirical approach. For this we assumed that the actual profile is symmetrical in respect to its central axis and that the point spread functions (PSFs) of both stars are the same. We constructed an empirical profile $\mathcal{P}_e$ using the side of the stellar profile of HAT-P-32A which is not distorted by the additional flux of HAT-P-32B and mirroring it at the central axis of the HAT-P-32A profile. We then used a scaled down, shifted in pixel position version of $\mathcal{P}_e$ to model the profile of HAT-P-32B. The final model for the superposition of both stellar profiles $\mathcal{W}$ was of the form:
\begin{equation}
\mathcal{W}\left(x\right)=
\mathcal{P}_e\left(x\right)+s\cdot\mathcal{P}_e\left(x-x_{A-B}\right)
\label{eq:mp}
\end{equation}
In this model the scaling factor ($s$), the center of the stellar profile of HAT-P-32A ($x$), and the distance of the centers of both stars ($x_{A-B}$) were free parameters. From the optimization of this model we obtained the flux for HAT-P-32A ($f_A$) and HAT-P-32B ($f_B$) for every wavelength element in every of the 25 images by summing over the respective theoretical profiles.

\subsubsection{Dilution of the optical transit depth measurements of HAT-P-32A b}
Since HAT-P-32B was within the aperture used in Sect. \ref{sec:specextract} to determine the flux of HAT-P-32A, its additional flux had a diluting effect on the transit depth. This effect is color dependent as HAT-P-32B contributes different amounts of flux in the individual wavelength channels defined in Sect. \ref{sec:lightcurves}. We determined the contribution in every wavelength channel by first summing the measured flux of both stars separately within the channel limits in each of the 25 frames. For each channel we then divided the resulting total flux of both stars and adopted the mean of the 25 flux ratios as the final result of the dilution factor  $c_n=f_A/f_B$ and their standard deviation as its uncertainty. The resulting values for each channel are listed in Table \ref{tab:tspec_results} in Sect. \ref{sec:results}.

\subsubsection{Resulting relative optical spectrum}
While for the correction of the multi-color channel light curves in this paper only low resolution information for HAT-P-32b was needed, higher resolution information was available. We used this data to determine the stellar properties of HAT-P-32B in Sect. \ref{sec:mdwarf2phoenix}. In order to remove the OSIRIS instrument sensitivity function from the data and clean it of any telluric absorption we used the relative spectrum of HAT-P-32B and HAT-P-32A for this analysis. Consequently, all wavelength-dependent telluric and instrumental effects are divided out. Since the slit alignment was optimized to center HAT-P-32A and the reference star (Ref2, see Table \ref{tab:coord}) within it, HAT-P-32B was not perfectly centered in the slit. As a consequence the wavelength solution for its spectrum is  slightly shifted compared to the wavelength solution of HAT-P-32A. We first applied this shift in wavelength and then calculated the flux ratio for every wavelength element in each of the 25 frames. We again adopted the mean of the 25 results as the final value and their standard deviation as the uncertainty.  The resulting relative spectrum covering the wavelength range $518-918$ nm is shown in Fig. \ref{fig:h32bvsa_optical}.

\subsection{Infrared photometry - WHT/LIRIS}
\begin{figure}
  \resizebox{\hsize}{!}{\includegraphics{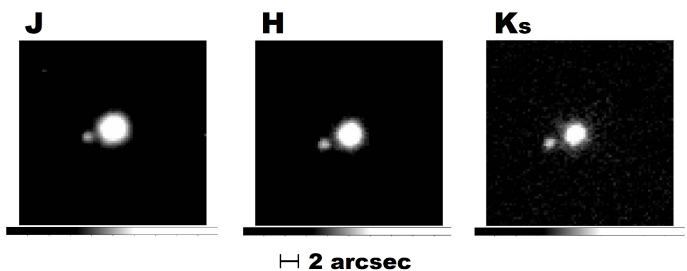}}
  	\caption{$J$, $H$, $K_S$ band photometry of HAT-P-32B and HAT-P-32A obtained with LIRIS at the WHT.}
	\label{fig:wht_phot}
\end{figure}

In addition to the optical spectrum we obtained $J$, $H$, and $K_S$ band measurements with the William Herschel Telescope (WHT) in the night of the 6th of October 2012 using LIRIS (Long-slit Intermediate Resolution Infrared Spectrograph) in imaging mode. We took 18 images and obtained 15 flats and darks 15 for each filter. A sample image of these observations for each filter is shown in Fig. \ref{fig:wht_phot}. We used PSF fitting to obtain the absolute flux for HAT-P-32B and HAT-P-32A. We then calculated the flux ratio $f_{\text{B}}/f_{\text{A}}$ for every image and adopted the mean as the final value and the standard deviation as the uncertainty for each filter. The results are listed in the first column of Table \ref{tab:c_ir}.

\defcitealias{2015ApJ...800..138N}{Ngo15}
\defcitealias{2013AJ....146....9A}{Adams13}
\defcitealias{2014ApJ...796..115Z}{Zhao14}
\begin{table}
\begin{center}
\begin{tabular}{c c c c c}
\hline\hline
$f_{\text{B}}/f_{\text{A}}$ [\%]&This work& \citetalias{2014ApJ...796..115Z}&\citetalias{2015ApJ...800..138N}\tablefootmark{a}
& \citetalias{2013AJ....146....9A}\tablefootmark{a}\\ 
 in Filter&&&&\\\hline 
$J$&$ 2.65\pm  0.45 $&...&\tol{2.19}{+0.57}{-0.56}& ... \\
$H$& $3.73\pm 0.65 $&$4.4\pm0.5$&\tol{3.41}{+0.23}{-0.22}& ...\\
$K'$&... & ...&\tol{3.21}{+0.13}{-0.12}& ...\\
$K_S$&$4.64 \pm 0.21$&$4.7\pm 0.2$&\tol{3.98}{+0.28}{-0.26}& 0.044\\
 \hline
\end{tabular}
\caption{Flux ratios between HAT-P-32B and HAT-P-32A obtained from infrared imaging in this work, \citetalias{2015ApJ...800..138N} \citep{2015ApJ...800..138N},  \citetalias{2014ApJ...796..115Z} \citep{2014ApJ...796..115Z} and  \citetalias{2013AJ....146....9A} \citep{2013AJ....146....9A}.\newline
{\tablefoottext{a}{Value calculated from the relative magnitudes $\Delta$mag given in the paper.}}}
\label{tab:c_ir}  
\end{center}
\end{table}

\subsection{Comparison to literature}\label{sec:mdwarf2lit}
\begin{table*}
\begin{center}
\begin{tabular}{c c c c c}
\hline\hline
Source&$T_\mathrm{eff}$ (K)& $\log g$ (cgs)&[Fe/H] (fixed/varied within prior)\\\hline
This work (2012-Sep-30)&\tol{3187}{+60}{-71}& \tol{4.94}{+0.50}{-0.68}&\tol{-0.04}{+0.08}{-0.08}\\
\citet{2014ApJ...796..115Z} (2013-Mar-02)&$3565 \pm82$&...&$-0.04$\\
 \citet{2015ApJ...800..138N}~ (2012-Feb-02)&$3516\pm12$&$4.8930\pm0.0098$&$0.00$\\
 \citet{2015ApJ...800..138N}~ (2013-Mar-02)&$3551 \pm 10$&$4.8677\pm0.0070$&$0.00$ \\\hline
\end{tabular}
\caption{Comparison of derived stellar parameters for HAT-P-32B.}  
\label{tab:stellarh32B}  
\end{center}
\end{table*}
We compared our results of the flux ratio $c_{\mathrm{filter}}=f_A/f_B$ of HAT-P-32B and HAT-P-32A with the broad band measurements from the literature obtained in the optical ($g'$, $r'$, $i'$, $z'$) by \citet{2014ApJ...796..115Z} with the Robo-AO instrument \citep{2014ApJ...790L...8B} on the 60 inch telescope at the Palomar Observatory and in the near infrared ($J$, $H$, $K'$ and $Ks$) by \citet{2014ApJ...796..115Z} and \citet{2015ApJ...800..138N} with NIRC2 a near-infrared imager (PI: Keith Matthews) using the AO system of Keck-II  \citep{2000PASP..112..315W} and \citet{2013AJ....146....9A} using MMT/Aries.\newline
The optical wavelength region probed with our GTC/OSIRIS transit observations fully overlaps with the $r'$ and $i'$ bands. We folded our data with the respective filter curves and found the results ($c_{r'}=0.0023(7)$, $c_{i'}=0.0064(8)$) to be consistent with the study of \citet{2014ApJ...796..115Z} ($c_{\textrm{Zhao},r'}=0.003(1)$, $c_{\textrm{Zhao},i'}=0.006(2)$).\newline
Since the WHT/LIRIS filter curves for the near infrared broadband filters $J$, $H$, and $K_S$ differ slightly from the ones used by the Keck-II/NIRC2 facility, an exact comparison between the respective measurements is not feasible. When neglecting these small differences in filter transmission, we, however, found that our results are consistent within $1\sigma$ with the near infrared values derived by all three studies  \citet{2014ApJ...796..115Z}, \citet{2015ApJ...800..138N} and \citet{2013AJ....146....9A}. Both \citet{2014ApJ...796..115Z} and \citet{2015ApJ...800..138N} independently analyzed the Keck-II/NIRC2 data obtained in the $H$ and $K_S$ band passes (while the $J$ and $K'$ band data was only analyzed by \citet{2015ApJ...800..138N}) and arrived at different results, which are only consistent with each other within $2\sigma$.

\subsection{Comparison to theoretical models}\label{sec:mdwarf2phoenix} 
\begin{figure*}
\resizebox{\hsize}{!}{\includegraphics{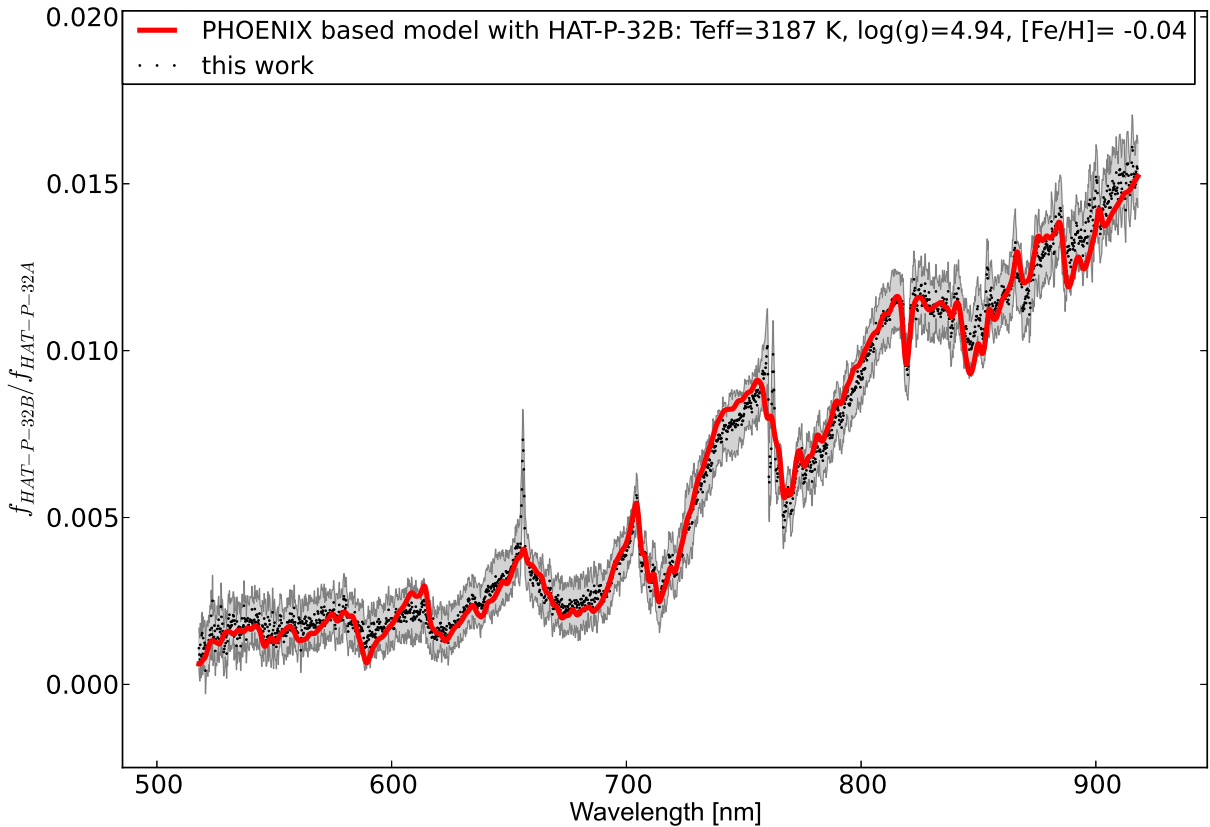}}
  	\caption{Relative optical spectrum HAT-P-32B/HAT-P-32A. The data is plotted as points, the $1\sigma$ uncertainty intervals are given as a light grey shaded area. The best fit PHOENIX model relative spectrum is given in red and relies on fixing the stellar properties of HAT-P-32A on literature values.}
	\label{fig:h32bvsa_optical}
\end{figure*}

\begin{figure*}
\resizebox{\hsize}{!}{\includegraphics{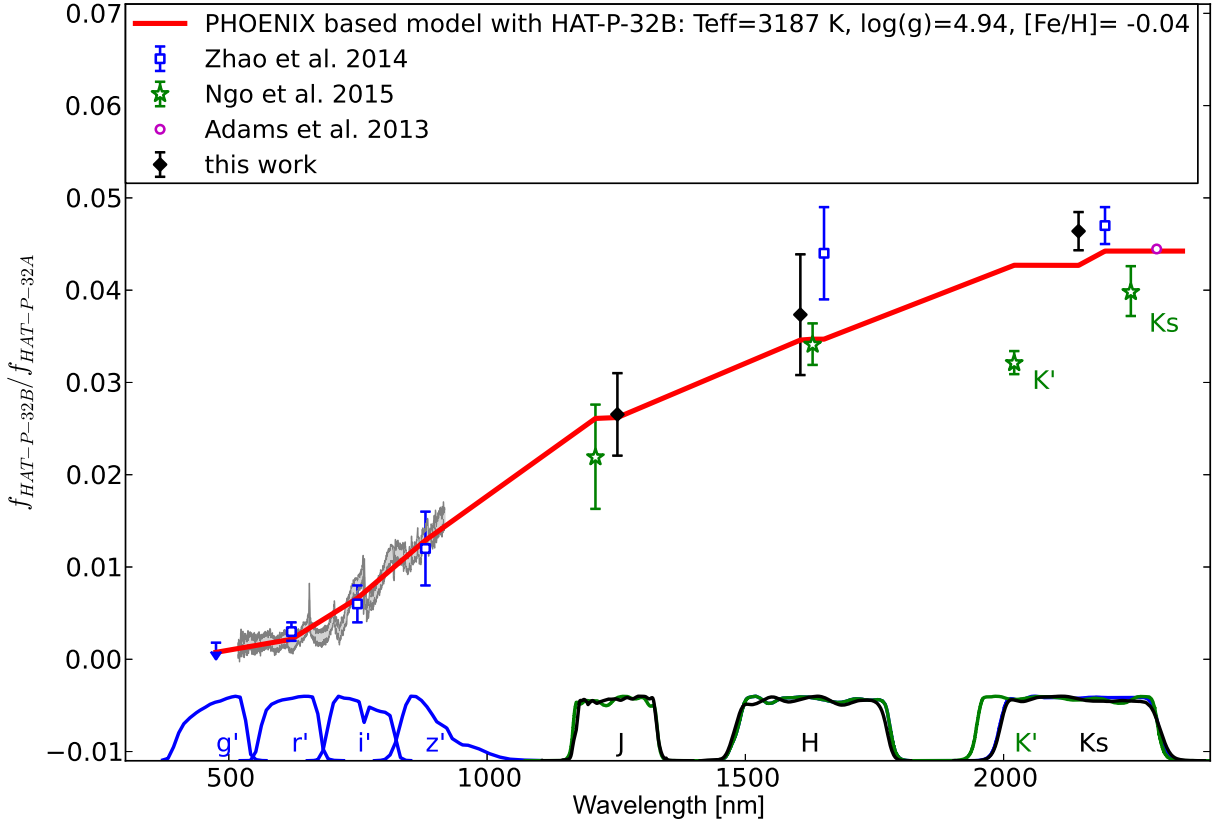}}
  	\caption{Relative optical and infrared photometry HAT-P-32B/HAT-P-32A. The results of the tree different literature studies are plotted together with our own $J$, $H$ and $K_S$ band measurements. The optical relative spectrum from Fig. \ref{fig:h32bvsa_optical} is also shown in light grey to emphasize the good agreement between this result and the literature photometry. The best fit PHOENIX model relative spectrum (same as in Fig. \ref{fig:h32bvsa_optical}) evaluated in the same band passes as the observations is plotted in red. The points obtained in the same or very similar band passes are plotted at slightly shifted abscissa values from the actual center of the respective band pass in order to increase clarity of the figure.}
	\label{fig:h32bvsa_ir}
\end{figure*}

Using their broadband measurements \citet{2014ApJ...796..115Z} and \citet{2015ApJ...800..138N} both determined physical properties of HAT-P-32B and arrived at similar results with effective temperatures around $T_\mathrm{eff}=3550$ K. The exact results are listed in Table \ref{tab:stellarh32B}.\newline
Both studies made use of the PHOENIX stellar atmosphere models but relied on slightly different approaches, stellar parameters for HAT-P-32A and different fixed values for the metallicity.\newline
We conducted our own analysis by fitting a theoretical model of the form:
\begin{equation}
\mathcal{R}=k\cdot \frac{M_{\mathrm{HAT-P-32B}}\left(\lambda,T_\mathrm{eff},\log g, \left[\mathrm{Fe/H}\right]\right)}{M_{\mathrm{HAT-P-32A}}\left(\lambda,T_\mathrm{eff},\log g, \left[\mathrm{Fe/H}\right]\right)} \label{eq:pspecmodel}
\end{equation}
to our optical and near infrared data as well as the literature broadband measurements (excluding the upper limit \citet{2014ApJ...796..115Z} give for the $g'$ band and the $K_S$ band data point of \citet{2013AJ....146....9A} for which no uncertainties were given).
In Eq. \ref{eq:pspecmodel} $k$ is a scaling factor (corresponding to the squared radius ratio of the two stars $\left(R_B/R_A\right)^2$) and $M_{\mathrm{HAT-P-32B}}$ and $M_{\mathrm{HAT-P-32A}}$ are PHOENIX stellar models, interpolated to specific stellar parameters from the model grid provided by \citet{2013A&A...553A...6H} with trilinear interpolation. During the fit the stellar parameters of HAT-P-32A were allowed to vary within the uncertainties of the given literature values ([Fe/H] $=-0.04\pm0.08$, $\log g$ (cgs) $=4.33 \pm 0.01$, \citet{2011ApJ...742...59H}, and $T_{\mathrm{eff}}=6269\pm64$ K, \citet{2014ApJ...796..115Z}) and an identical metallicity for both stars was enforced. For the comparison with the broadband points we folded the PHOENIX model with the respective filter curves. In this step we used the exact instrument specific filter curve (downloaded from the facility web-pages) for each broadband data point. For the optical spectrum we folded the PHOENIX model spectrum with a Gauss function to reduce the resolution to match the data.\newline 
We found that if we only fit the broadband measurements we arrive at similar values for HAT-P-32B's stellar parameters as the ones derived by \citet{2014ApJ...796..115Z} and \citet{2015ApJ...800..138N}. When including the optical data, however, the fit clearly favored cooler temperatures. Due to the inconsistencies in the literature regarding the results obtained for the $H$ and $K_S$ bands with Keck-II/NIRC2 we decided to only include our own data in the final optimization.  The difference in the best fit effective temperature due to this exclusion of broadband points is insignificant (6 K). Our results for HAT-P-32B's stellar parameters are given in Table \ref{tab:stellarh32B}. The uncertainties of these results were inflated with the red noise factor $\beta$, which was calculated as described in Sect. \ref{sec:rednoise}. The errors do not incorporate any intrinsic uncertainties of the PHOENIX stellar atmosphere models.\newline
The best fit model relative spectrum is plotted together with all data points (including those that were not regarded in the fit) in Fig. \ref{fig:h32bvsa_optical} and \ref{fig:h32bvsa_ir}.\newline
It stands out that the $K'$ band result of \citet{2015ApJ...800..138N} diverges significantly from all $K_S$ band values including their own. This offset is too large to be explained by the difference between the $K'$ and $K_S$ band passes. Since the $K_S$ and $K'$ observations by \citet{2015ApJ...800..138N} were conducted at different dates (13 month apart), a more likely explanation for this inconsistency could be stellar variability of HAT-P-32B due to activity i.e. star spots or flares. We measure a prominent $H_{\alpha}$ emission line at $656$ nm in the optical spectrum (see Fig. \ref{fig:h32bvsa_optical}), indicative of such stellar activity.

\end{appendix}
\bibliographystyle{aa} 
\bibliography{paper_h32_arXiv}
\end{document}